%% file: main.tex
\documentclass[letterpaper,twocolumn,10pt]{article}
\usepackage{template/usenix-2020-09}

\input{custom_style}

\usepackage[english]{babel}
\newcounter{bug}
\renewcommand*{\thebug}{\textbf{V\arabic{bug}}}

\usepackage[english]{babel}
\newenvironment{bug_ns}[1][]{\refstepcounter{bug}
\noindent\textbf{Vulnerability \thebug#1}\rmfamily}

\newcounter{challenge}
\renewcommand*{\thechallenge}{\textbf{C\arabic{challenge}}}
\newenvironment{chal}[1][]{\refstepcounter{challenge}
\noindent\textbf{\thechallenge#1.}\rmfamily}

\newcounter{observation}
\renewcommand*{\theobservation}{\textbf{O\arabic{observation}}}
\newenvironment{obs}[1][]{\refstepcounter{observation}
\noindent\textbf{Observation \theobservation#1}\rmfamily}

\begin{document}

\date{}

\title{\Large \bf HyPFuzz: Formal-Assisted Processor Fuzzing}

\author{
{\rm Chen Chen$^\dagger$, Rahul Kande$^\dagger$, Nathan Nguyen$^\dagger$, Flemming Andersen$^\dagger$, Aakash Tyagi$^\dagger$,}\\ 
{\rm Ahmad-Reza Sadeghi$^\ast$, and Jeyavijayan Rajendran$^\dagger$}\\
$^\dagger$Texas A\&M University, USA, $^\ast$Technische Universit\"at Darmstadt, Germany\\
{\tt $^\dagger$\{chenc, rahulkande, nathan.tm.nguyen, flandersen, tyagi, jv.rajendran\}@tamu.edu,}\\
{\tt $^\ast$\{ahmad.sadeghi\}@trust.tu-darmstadt.de}
} 

\maketitle

\input{openarch/abstract}

\input{openarch/introduction}
\input{openarch/background}
\input{openarch/motivation}
\input{openarch/method}

\input{openarch/experiment}
\input{openarch/relatedwork}
\input{openarch/discussion}
\input{openarch/Conclusion}

\bibliographystyle{plain_auth_limited}
\bibliography{Bibfile.bib}

\input{openarch/Appendix}
\end{document}

%% file: custom_style.tex
\usepackage{tikz}
\usepackage{amsmath}
\usepackage{array}
\usepackage{comment}
\usepackage{filecontents}
\usepackage{tcolorbox}
\usepackage{listings}
\input{listing_style}
\usepackage[ruled,vlined,linesnumbered]{algorithm2e}
\usepackage[normalem]{ulem}  

\usepackage{caption}
\usepackage{subcaption}
\usepackage{xurl}
\usepackage{bold-extra}
\usepackage{courier}

\usepackage{bm} 
\usepackage{soul} 

\newcommand{\ourtool}{\textit{HyPFuzz}}
\newcommand{\thehuzz}{\textit{TheHuzz}}
\newcommand{\rfuzz}{\textit{RFUZZ}}
\newcommand{\difuzz}{\textit{DIFUZZRTL}}

\newcommand{\rc}{{\tt Rocket Core}}
\newcommand{\boom}{{\tt BOOM}}
\newcommand{\cva}{{\tt CVA6}}
\newcommand{\morkx}{{\tt mor1kx}}
\newcommand{\orth}{{\tt OR1200}}

\newcommand{\systemverilog}{SystemVerilog}
\newcommand{\verilog}{Verilog}
\newcommand{\chisel}{Chisel}
\newcommand{\JG}{\textit{JasperGold}}
\newcommand{\vcs}{\textit{VCS}}
\newcommand{\verilator}{\textit{Verilator}}
\newcommand{\modelsim}{\textit{Modelsim}}
\newcommand{\vcformal}{\textit{VC Formal}}
\newcommand{\questa}{\textit{Questa}}

\newcommand{\riscv}{RISC-V}

\newcommand{\spike}{\textit{Spike}}

\newcommand{\maxPerMoreTh}{6.84} 
\newcommand{\avgPerMoreThe}{1.72} 
\newcommand{\avgPerMoreRR}{4.73}  

\newcommand{\minDiffCovStrats}{0.5}
\newcommand{\cvaDiffCovStrats}{1.31}

\newcommand{\covPerMoreOrthHyPTheH}{0.40}

\newcommand{\aveCovSpeedup}{11.68} 
\newcommand{\ratecva}{41.24} 
\newcommand{\rateboom}{11.42}  

\newcommand{\minSpeedupThreeprocs}{1.5} 

\newcommand{\aveCovSpeedupvsRandreg}{239.93} 

\newcommand{\avebugtimerate}{3.06} 
\newcommand{\avebuginstrate}{3.05} 

\newcommand{\CondAveCovImprv}{15.95} 
\newcommand{\FSMAveCovImprv}{33.76} 

\newcommand{\randsel}{\textit{RandSel}}
\newcommand{\maxuncov}{\textit{MaxUncovd}}
\newcommand{\moddep}{\textit{ModDep}}
\newcommand{\bottop}{\textit{BotTop}}

\usepackage{enumitem,amssymb}
\newlist{todolist}{itemize}{2}
\setlist[todolist]{label=$\square$}

\def\todoen{1}
\def\rahulen{0}
\def\chenen{0}
\def\nathanen{1}
\def\removesen{0}
\def\adden{0}

\usepackage{graphicx}

\usepackage{diagbox}

\usepackage{multirow}
\usepackage[symbol]{footmisc}

\if 1\todoen 
\newcommand\todo[1]{\textcolor{red}{\textbf{TODO:} #1}}
\else
\newcommand\todo[1]{}
\fi

\if 1\rahulen

\else

\fi

\if 1\chenen
\newcommand\ch[1]{\textcolor{blue}{ #1}}
\else
\newcommand\ch[1]{#1}
\fi

\if 1\nathanen

\else

\fi

\if 1\removesen
\newcommand\red[1]{\textcolor{red!40!white}{\sout{#1}}}
\else
\newcommand\red[1]{}
\fi

\if 1\adden
\newcommand\blue[1]{\textcolor{blue}{#1}}
\else
\newcommand\blue[1]{{#1}}
\fi

\usepackage{tikz}
\newcommand{\tikzxmark}{%
\tikz[scale=0.23] {
    \draw[red,line width=0.7,line cap=round] (0,0) to [bend left=6] (1,1);
    \draw[red,line width=0.7,line cap=round] (0.2,0.95) to [bend right=3] (0.8,0.05);
}}
\newcommand{\tikzcmark}{%
\tikz[scale=0.23] {
    \draw[blue,line width=0.7,line cap=round] (0.25,0) to [bend left=10] (1,1);
    \draw[blue,line width=0.8,line cap=round] (0,0.35) to [bend right=1] (0.23,0);
}}

\usepackage{subcaption} 


%% file: listing_style.tex
\definecolor{codegreen}{rgb}{0,0.6,0}
\definecolor{codegray}{rgb}{0.5,0.5,0.5}
\definecolor{codepurple}{rgb}{0.58,0,0.82}
\definecolor{backcolour}{rgb}{0.95,0.95,0.92}

\let\othelstnumber=\thelstnumber
\def\createlinenumber#1#2{
    \edef\thelstnumber{%
        \unexpanded{%
            \ifnum#1=\value{lstnumber}\relax
              #2%
            \else}%
        \expandafter\unexpanded\expandafter{\thelstnumber\othelstnumber\fi}%
    }
    \ifx\othelstnumber=\relax\else
      \let\othelstnumber\relax
    \fi
}

\lstdefinestyle{customc}{
  belowcaptionskip=1\baselineskip,
  breaklines=true,
  frame=single,
  xleftmargin=0.35cm,
  xrightmargin=0.15cm,
  numbers=left,
  numbersep=5pt,  
  language=C,
  showstringspaces=false,
  basicstyle=\footnotesize\ttfamily,
  keywordstyle=\bfseries\color{green!40!black},
  commentstyle=\itshape\color{purple!40!black},
  identifierstyle=\color{blue},
  stringstyle=\color{orange},
}

\lstdefinestyle{customcArianeExploit1}{
  breaklines=true,
  frame=single,
  xleftmargin=0.4cm,
  xrightmargin=0.2cm,
  numbers=left,
  numbersep=5pt,  
  language=C,
  showstringspaces=false,
  basicstyle=\footnotesize\ttfamily,
  keywordstyle=\bfseries\color{green!40!black},
  commentstyle=\itshape\color{purple!60!black},
  identifierstyle=\color{blue},
  stringstyle=\color{yellow!50!black},
  morekeywords={asm},
  keywordstyle=[2]\bfseries\color{brown!60!black},
}

\lstdefinestyle{customcArianeExploit}{
  breaklines=true,
  frame=single,
  xleftmargin=0.4cm,
  xrightmargin=0.2cm,
  numbers=left,
  numbersep=5pt,  
  language=C,
  showstringspaces=false,
  basicstyle=\footnotesize\ttfamily,
  keywordstyle=\bfseries\color{blue},
  commentstyle=\itshape\color{green!50!black},
  identifierstyle=\color{black},
  stringstyle=\color{brown},
  morekeywords={asm},
  keywordstyle=[2]\bfseries\color{black},
}

\lstdefinestyle{customlog}{
  breaklines=true,
  frame=single,
  xleftmargin=0.35cm,
  xrightmargin=0.15cm,
  numbers=left,
  numbersep=5pt,  
  language=C,
  showstringspaces=false,
  basicstyle=\footnotesize\ttfamily,
  keywordstyle=\color{blue},
  commentstyle=\itshape\color{purple!40!black},
  identifierstyle=\color{blue},
  stringstyle=\color{orange},
  keywords=[2]{INFO},
  keywords=[3]{ERROR},x
  keywordstyle=[2]\bfseries\color{green!40!black},
  keywordstyle=[3]\bfseries\color{red!500!black},
}

\definecolor{verilogcommentcolor}{RGB}{104,180,104}
\definecolor{verilogkeywordcolor}{RGB}{49,49,255}
\definecolor{verilogsystemcolor}{RGB}{128,0,255}
\definecolor{verilognumbercolor}{RGB}{255,143,102}
\definecolor{verilogstringcolor}{RGB}{160,160,160}
\definecolor{verilogdefinecolor}{RGB}{128,64,0}
\definecolor{verilogoperatorcolor}{RGB}{0,0,128}
\definecolor{pointcolor}{RGB}{192,0,0} 
\lstdefinestyle{prettyverilog}{
   language           = Verilog,
   commentstyle       = \color{verilogcommentcolor},
   alsoletter         = \$'0123456789\`,
   literate           = *{+}{{\verilogColorOperator{+}}}{1}%
                         {-}{{\verilogColorOperator{-}}}{1}%
                         {@}{{\verilogColorOperator{@}}}{1}%
                         {;}{{\verilogColorOperator{;}}}{1}%
                         {*}{{\verilogColorOperator{*}}}{1}%
                         {?}{{\verilogColorOperator{? }}}{1}%
                         {:}{{\verilogColorOperator{:}}}{1}%
                         {<}{{\verilogColorOperator{<}}}{1}%
                         {>}{{\verilogColorOperator{>}}}{1}%
                         {!}{{\verilogColorOperator{!}}}{1}%
                         {^}{{\verilogColorOperator{^}}}{1}%
                         {|}{{\verilogColorOperator{| }}}{1}%
                         {||}{{\verilogColorOperator{|| }}}{1}%
                         {=}{{\verilogColorOperator{= }}}{1}%
                         {==}{{\verilogColorOperator{== }}}{1}%
                         {=>}{{\verilogColorOperator{=> }}}{1}%
                         {[}{{\verilogColorOperator{[}}}{1}%
                         {]}{{\verilogColorOperator{]}}}{1}%
                         {(}{{\verilogColorOperator{(}}}{1}%
                         {)}{{\verilogColorOperator{)}}}{1}%
                         {,}{{\verilogColorOperator{,}}}{1}%
                         {.}{{\verilogColorOperator{.}}}{1}%
                         {~}{{\verilogColorOperator{$\sim$}}}{1}%
                         {\%}{{\verilogColorOperator{\%}}}{1}%
                         {\&}{{\verilogColorOperator{\& }}}{1}%
                         {\&\&}{{\verilogColorOperator{\&\& }}}{1}%
                         {\#}{{\verilogColorOperator{\#}}}{1}%
                         {\ /\ }{{\verilogColorOperator{\ /\ }}}{3}%
                         {\ _}{\ \_}{2}%
                        ,
   morestring         = [s][\color{verilogstringcolor}]{"}{"},%
   identifierstyle    = \color{black},
   vlogdefinestyle    = \color{verilogdefinecolor},
   vlogconstantstyle  = \color{verilognumbercolor},
   vlogsystemstyle    = \color{verilogsystemcolor},
   basicstyle         = \scriptsize\fontencoding{T1}\ttfamily,
  columns=fullflexible, 
   keywordstyle       = \bfseries\color{verilogkeywordcolor},
   morekeywords      = {val, when, port, coverage, unique},
   numbers            = left,
   numbersep          = 5pt,
   tabsize            = 2,
   escapeinside       = {/*!}{!*/},
   upquote            = true,
   sensitive          = true,
   showstringspaces   = false, 
   frame              = single,
   breaklines         = true,
   abovecaptionskip   = 0pt,
   belowcaptionskip   = 2pt,   
   xleftmargin        =0.35cm,
   xrightmargin       =0.15cm,
   captionpos         = t,
   emph               = {Point, Point0, Point1, Point2, Point3, Point4, Point5, Point6, Point7, Point8, Point9},
   emphstyle          =\color{pointcolor},
   emph               = {[2] STVEC,SCOUNTEREN,MSTATUS,MTVEC,ML1_ICACHE_MISS,ML1_DCACHE_MISS,MITLB_MISS,MDTLB_MISS,
                             MLOAD,MSTORE,MEXCEPTION,MEXCEPTION_RET,MBRANCH_JUMP,MCALL,MRET,MMIS_PREDICT,MSB_FULL,
                             MIF_EMPTY,MHPM_COUNTER_17,MHPM_COUNTER_18,MHPM_COUNTER_19,MHPM_COUNTER_20,MHPM_COUNTER_21,
                             MHPM_COUNTER_22,MHPM_COUNTER_23,MHPM_COUNTER_24,MHPM_COUNTER_25,MHPM_COUNTER_26,MHPM_COUNTER_27,
                             MHPM_COUNTER_28,MHPM_COUNTER_29,MHPM_COUNTER_30,MHPM_COUNTER_31}, 
   emphstyle          = {[2]\bfseries\color{verilogkeywordcolor}}
}

\makeatletter

\newcommand\language@verilog{Verilog}
\expandafter\lst@NormedDef\expandafter\languageNormedDefd@verilog%
  \expandafter{\language@verilog}
  
\lst@SaveOutputDef{`'}\quotesngl@verilog
\lst@SaveOutputDef{``}\backtick@verilog
\lst@SaveOutputDef{`\$}\dollar@verilog

\newcommand\getfirstchar@verilog{}
\newcommand\getfirstchar@@verilog{}
\newcommand\firstchar@verilog{}
\def\getfirstchar@verilog#1{\getfirstchar@@verilog#1\relax}
\def\getfirstchar@@verilog#1#2\relax{\def\firstchar@verilog{#1}}

\newcommand\addedToOutput@verilog{}
\lst@AddToHook{Output}{\addedToOutput@verilog}

\newcommand\constantstyle@verilog{}
\lst@Key{vlogconstantstyle}\relax%
   {\def\constantstyle@verilog{#1}}

\newcommand\definestyle@verilog{}
\lst@Key{vlogdefinestyle}\relax%
   {\def\definestyle@verilog{#1}}

\newcommand\systemstyle@verilog{}
\lst@Key{vlogsystemstyle}\relax%
   {\def\systemstyle@verilog{#1}}

\newcount\currentchar@verilog
  
\newcommand\@ddedToOutput@verilog
{%
   \ifnum\lst@mode=\lst@Pmode%
      \expandafter\getfirstchar@verilog\expandafter{\the\lst@token}%
      \expandafter\ifx\firstchar@verilog\backtick@verilog
         \let\lst@thestyle\definestyle@verilog%
      \else
         \expandafter\ifx\firstchar@verilog\dollar@verilog
            \let\lst@thestyle\systemstyle@verilog%
         \else
            \expandafter\ifx\firstchar@verilog\quotesngl@verilog
               \let\lst@thestyle\constantstyle@verilog%
            \else
               \currentchar@verilog=48
               \loop
                  \expandafter\ifnum%
                  \expandafter`\firstchar@verilog=\currentchar@verilog%
                     \let\lst@thestyle\constantstyle@verilog%
                     \let\iterate\relax%
                  \fi
                  \advance\currentchar@verilog by \@ne%
                  \unless\ifnum\currentchar@verilog>57%
               \repeat%
            \fi
         \fi
      \fi
   \fi
}

\lst@AddToHook{PreInit}{%
  \ifx\lst@language\languageNormedDefd@verilog%
    \let\addedToOutput@verilog\@ddedToOutput@verilog%
  \fi
}

\newcommand{\verilogColorOperator}[1]
{%
  \ifnum\lst@mode=\lst@Pmode\relax%
   {\bfseries\textcolor{verilogoperatorcolor}{#1}}%
  \else
    #1%
  \fi
}

\makeatother

\lstdefinestyle{mystyle}{
    commentstyle=\textit,
    keywordstyle=\textbf,
    stringstyle=\color{codepurple},
    basicstyle=\ttfamily,
    breakatwhitespace=false,         
    breaklines=true,      
    frame=single, 
    framexleftmargin=\parindent,
    captionpos=b,                    
    keepspaces=true,                 
    numbers=left,    
    numberstyle=\normalsize,
    stepnumber=1,
    numbersep=5pt,   
    xleftmargin=1.5\parindent,
    showspaces=false,                
    showstringspaces=false,
    showtabs=false,                  
    tabsize=2
}

%% file: openarch/abstract.tex
\begin{abstract}
Recent research has shown that hardware fuzzers can effectively detect security vulnerabilities in modern processors.
However, existing hardware fuzzers do not fuzz well the hard-to-reach design spaces.
Consequently, these fuzzers cannot effectively fuzz security-critical control- and data-flow logic in the processors, hence missing security vulnerabilities.

To tackle this challenge, we present \ourtool{}, a hybrid fuzzer that leverages formal verification tools to help fuzz the hard-to-reach part of the processors. 
To increase the effectiveness of \ourtool{}, we perform optimizations in time and space.
First, we develop a scheduling strategy to \blue{prevent under- or over-utilization of the capabilities of formal tools and fuzzers.}
Second, we develop heuristic strategies to select points in the design space for the formal tool to target.

We evaluate \ourtool{} on five widely-used open-source processors. 
\blue{\ourtool{} detected} all the vulnerabilities detected by the most recent processor fuzzer and found three new vulnerabilities that \blue{were missed by previous extensive fuzzing and formal verification.} 
This led to two new common vulnerabilities and exposures~(CVE) entries.
\ourtool{} also achieves 
\aveCovSpeedup{}$\times$ faster coverage than the most recent processor fuzzer.
\end{abstract} 






%% file: openarch/introduction.tex
\section{Introduction}\label{sec:intro}
Hardware designs are becoming increasingly complex to meet the rising need for custom hardware and increased performance. 
Around 67\% of the application-specific integrated circuit~(ASIC) designs developed in 2020 have over 1 \blue{million} gates, and 45\% of them embed two or more processors~\cite{wilsonstudy}.
However, unlike software vulnerabilities that can be patched, most hardware vulnerabilities cannot be fixed post-fabrication, resulting in security vulnerabilities that put many critical systems at risk and tarnish the reputation of the companies involved.
Hence, it is essential to detect vulnerabilities pre-fabrication. However, the emergence of hardware security vulnerabilities~\cite{dessouky2019hardfails, Lipp2018meltdown, Kocher2018spectre}
show\blue{s} that \red{they}\blue{vulnerabilities} are becoming more stealthy and harder to detect~\cite{dessouky2019hardfails}. 
MITRE reports 111 hardware-related common weakness enumerations~(CWEs) as of 2022~\cite{MITRE}. There exist a variety of traditional\red{ and trending} methodologies and tools for hardware security verification, each having its own advantages and shortcomings, as we explain below.

\noindent \textbf{Hardware verification techniques.}
Academic and industry researchers have developed numerous hardware vulnerability detection techniques. These techniques can be classified as (i)~formal: theorem proving~\cite{cyrluk1994effective}, formal assertion proving~\cite{witharana2022survey}, model checking~\cite{clarke2018handbook}, and information-flow tracking~\cite{hu2021hardware}\blue{;} and (ii)~simulation-based: random regression~\cite{naveh2007constraint}
and hardware fuzzing~\cite{rfuzz,hur2021difuzzrtl,fuzzhwlikesw,kandethehuzz}.

Formal verification techniques \red{check}\blue{prove} whether a design-under-test~(DUT) satisfies specified properties~\cite{introformal}.
However, these techniques alone cannot verify the \red{entirety of the}\blue{entire} DUT because: 
(i)~\blue{in most cases,} writing properties requires manual effort and expert knowledge of the DUT\red{; this step}\blue{, which} is error-prone and time-consuming~\cite{kandethehuzz,dessouky2019hardfails}, and 
(ii)~large DUTs (such as processors) lead to state explosion, making it impractical to comprehensively verify a \red{processor}\blue{DUT} for security vulnerabilities~\cite{dessouky2019hardfails,clarke2001progress}. 

Random regression can automatically generate test cases for verification. However, it does not scale well to large DUTs, including processors~\cite{rfuzz, hur2021difuzzrtl, kandethehuzz}, especially the regions of the design that are hard-to-reach. 
For example, the probability of a random regression technique \blue{to generate a test case that triggers}\red{triggering} a zero flag (indicates that the output is ``0'') in a 64-bit subtract module is $2^{-\red{32}\blue{64}}$.
Unfortunately, many hardware security-critical components are inherently hard-to-reach (e.g., access control\red{,} \blue{and} password checkers)~\cite{dessouky2019hardfails}.


Inspired by the success of software fuzzing methods, researchers have investigated hardware fuzzing to significantly increase the \red{design space exploration}\blue{exploration of design spaces} and accelerate the detection of security vulnerabilities~\cite{rfuzz, hur2021difuzzrtl, kandethehuzz, fuzzhwlikesw}. 
Unfortunately, software fuzzers cannot be directly applied to the software model of hardware due to \blue{their} fundamental differences~\cite{fuzzhwlikesw, verilator, rfuzz, kandethehuzz}; for instance, hardware does not have an equivalent of a software crash, and software does not have floating wires. 
Hardware fuzzers outperform traditional hardware verification techniques, such as random regression and formal verification techniques~\cite{rfuzz, fuzzhwlikesw, hur2021difuzzrtl, kandethehuzz}, in terms of coverage, scalability, and efficiency in detecting vulnerabilities, and they can fuzz large designs such as processors~\cite{hur2021difuzzrtl, kandethehuzz}, including \rc{}~\cite{rocket_chip_generator} and \cva{}~\cite{cva6}~\cite{hur2021difuzzrtl}. They have found vulnerabilities that lead to privilege escalation and arbitrary code execution attacks~\cite{kandethehuzz}. 
To improve efficiency, these fuzzers use coverage data that succinctly captures different hardware behaviors---finite-state machines (FSMs), branch conditions, statements, multiplexors, etc.---to generate and mutate new test cases. 



However, while hardware fuzzing is very promising, it still does not even cover 70\% of the hardware design in a practical amount of time. 
For example, a recent hardware fuzzer, \thehuzz{}, which has \red{better}\blue{higher} and faster coverage 
than random regression techniques and \difuzz{}~\cite{hur2021difuzzrtl}, has covered only about $63\%$ of the total coverage points in the processors, leaving one-third of the space unexplored for vulnerabilities~\cite{kandethehuzz}.


The coverage of current hardware fuzzers falls well below industry standards. For instance, Google states that security-critical programs should \red{reach the highest level achieving at least 90\% coverage}\blue{achieve at least 90\% coverage}~\cite{ivankovic2019code}. 
\blue{Achieving 90\% coverage is also typical in hardware verification~\cite{wile2005comprehensive}. Faster coverage can promote the decision to tape out and expose unverified design spaces and vulnerabilities early~\cite{verifiwhitepaper}.}
\red{Hardware---often treated as the root of trust for many software security applications---should also be fuzzed until the coverage hits at least 90\%.}Unfortunately, none of the existing hardware \blue{processor} fuzzers meet these criteria. 

\begin{figure}
    \centering
    \includegraphics[trim=25 20 20 70,clip,width=0.75\linewidth]{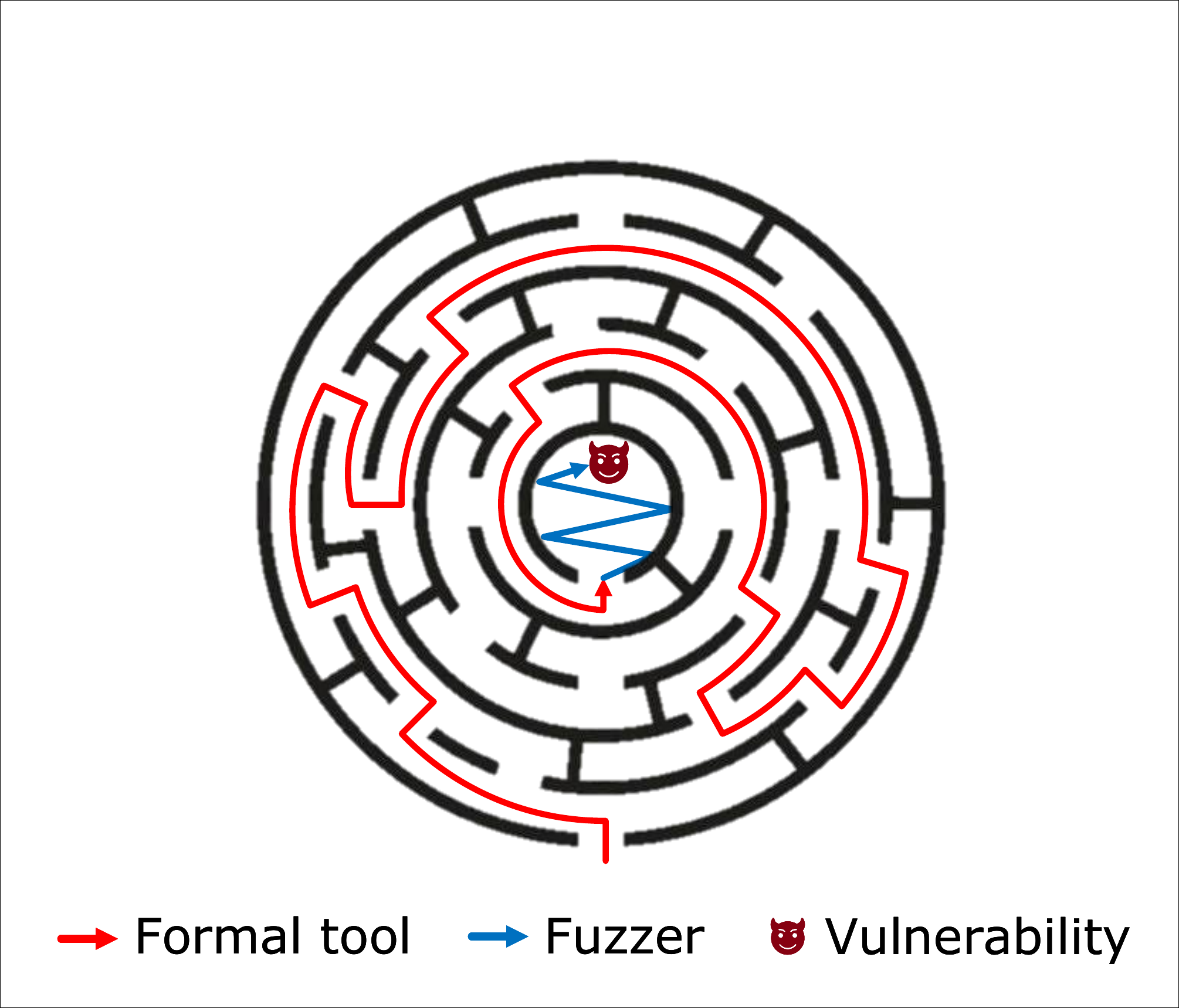}
    \caption{Formal tools and fuzzers catch vulnerabilities in the maze of designs.}
    \label{fig:high_level}
\end{figure}

\noindent \textbf{Our goals and contributions.}
To alleviate the above limitation of hardware fuzzers\red{,} and inspired by hybrid software fuzzers~\cite{stephens2016driller}, we aim at making the first step towards building a \emph{hybrid hardware fuzzer} that combines the capabilities of formal verification techniques/tools and fuzzing tools. 
Figure~\ref{fig:high_level} illustrates the intuition of a \textbf{hybrid hardware fuzzer}. The maze represents the entire design space of hardware, and the walls represent the conditions required to reach a design space. 
Formal tools will lead fuzzers to the hard-to-reach design spaces so that fuzzers can quickly explore them and detect the vulnerability in the hard-to-reach design spaces.

To this end, we developed a new hybrid hardware fuzzer, \ourtool{},
\red{We stress that hybrid software fuzzers cannot be utilized for hardware verification due to the differences between hardware and software properties mentioned above \cite{verilator, kandethehuzz, hur2021difuzzrtl}.}
\red{Building a hybrid hardware fuzzer}\blue{which} is non-trivial due to the following reasons. The first challenge is to build an \red{optimal}\blue{dynamic} time scheduling between the formal tool and the fuzzer since \red{non-optimal}\blue{static} scheduling leads to under- or over-utilization of the capabilities of the formal tool and the fuzzer.
The second challenge concerns the \red{optimal }selection of 
coverage points
in the DUT to be targeted by the formal tool and the fuzzer, as the former is better at reaching hard-to-reach design spaces while the latter is better at exploring design spaces. 
The third challenge concerns the incompatibility of formal tools and fuzzing tools:  
formal tools target assertions about properties of the hardware, whereas fuzzers target coverage points of the hardware. For a seamless integration of formal and fuzzing tools, one needs to convert these assertions into coverage points, and vice-versa.

To solve these challenges: 
(i)~We create a scheduling strategy for fuzzer and formal tool that \red{optimally }increases the overall coverage rate of \ourtool{} (see Section~\ref{sec:scheduler}). 
(ii)~We propose multiple strategies to select the coverage points for the formal tool and empirically determine the best-performing strategy (see Section~\ref{sec:selector}). 
(iii)~We develop a custom \textit{property generator} that converts coverage points to assertions \blue{taken by formal tools} and a custom \textit{test case converter} that converts Boolean assignments from formal tools into test cases taken by fuzzers, enabling seamless integration of fuzzing and formal techniques for hardware (see Section~\ref{sec:basic_framework}). 

Consequently, \ourtool{} achieves \aveCovSpeedup{}${\times}$ faster coverage than the most recently proposed processor fuzzer, \thehuzz{}~\cite{kandethehuzz}. 
It has detected three new vulnerabilities, apart from detecting all the vulnerabilities previously reported. It is also \avebugtimerate{}$\times$ faster than \thehuzz{}. 
 
In summary, our main contributions are:
\begin{itemize}[align=parleft,leftmargin=*]
    \item We present a novel processor fuzzer, \ourtool{}, which combines fuzzing and formal verification techniques to verify large-scale processor designs and supports commonly-used hardware description languages~(HDLs) like \verilog{} and \systemverilog{}. We use \red{optimal }scheduling and selection strategies making \ourtool{} fast and efficient for design space exploration and vulnerability detection. 
    
    \item We evaluate the effectiveness of \ourtool{} on five real-world open-source processors from RISC-V instruction set architecture (ISA)~\cite{riscv_home}---\rc{}~\cite{rocket_chip_generator}, \cva{}~\cite{cva6}, and \boom{}~\cite{boom}---and OpenRISC ISA~\cite{openrisc_home}---\morkx{}~\cite{mor1kx} and \orth{}~\cite{or1200}--- which are widely used as benchmarks in the hardware security community and include all the benchmarks used by \difuzz{}~\cite{hur2021difuzzrtl} and \thehuzz{}~\cite{kandethehuzz}. 

    \item \ourtool{} achieves \aveCovSpeedup{}${\times}$ faster coverage than the most recent processor fuzzer and \aveCovSpeedupvsRandreg{}${\times}$ faster coverage than random regression. \blue{It found three new vulnerabilities leading to two common vulnerabilities and exposures~(CVE) entries, \texttt{CVE-2022-33021} and \texttt{CVE-2022-33023}, apart from detecting all the vulnerabilities detected by \thehuzz{}. \ourtool{} is \avebugtimerate{}$\times$ faster regarding run-time and \avebuginstrate{}$\times$ faster regarding the number of instructions than \thehuzz{}. }
    
    
    \red{\item} \red{\ourtool{} has found three new vulnerabilities leading to two common vulnerabilities and  exposures~(CVE) entries,  \texttt{CVE-2022-33021} and \texttt{CVE-2022-33023}, apart from detecting all the vulnerabilities detected by \thehuzz{}. \ourtool{} is \avebugtimerate{}$\times$ faster than \thehuzz{} in terms of run-time and  \avebuginstrate{}$\times$ faster than \thehuzz{} in terms of the number of instructions. }
 
\end{itemize}

%% file: openarch/background.tex
\section{Background} \label{sec:background}
We now provide a succinct background on formal verification and hardware fuzzing, which form the basis of \ourtool{}.
 
\subsection{Formal Verification}\label{sec:formalVerif}
Formal verification techniques have shown to be effective at finding subtle vulnerabilities~\cite{clarke2018handbook},
such as side-channel leakage~\cite{manerkar2017rtlcheck,trippel2018checkmate,fadiheh2020formal,ponce2022cats}, information leakage~\cite{deutschbein2022toward,rajendran2016formal,eldib2014formal}, and concurrency errors~\cite{das2020formal,graja2020comprehensive}. 
These techniques can \red{successfully }find these vulnerabilities because they can exhaustively \red{verify}\blue{prove} whether a design-under-test~(DUT) satisfies \red{a set of }specified properties. 

\blue{Existing hybrid software fuzzers use symbolic execution to generate test cases and explore the hard-to-reach regions of a DUT~\cite{zhu2022fuzzing,godefroid2005dart,godefroid2008automated,pak2012hybrid,pham2016model,zhao2019send,stephens2016driller}. 
This is done by identifying the execution paths a fuzzer cannot reach and generating test cases that force the target DUT to execute these paths~\cite{godefroid2005dart}. Symbolic execution explores all possible execution paths of a DUT, but it requires a mapping between other coverage metrics to execution paths and is limited by the huge space of execution path of large designs~\cite{zhu2022fuzzing}.}

\blue{\ourtool{} uses another method as it uses commercial hardware formal tools, like Cadence \JG{}~\cite{jaspergold}, to generate such test cases. \JG{} requires \systemverilog{} Assertion~(SVA) properties~\cite{IEEEstd} known as {\em cover} properties as inputs to enable \ourtool{} to generate test cases.} 
The properties \red{checked}\blue{proved} by such tools mostly fall under two categories: 
(i)~``assert/safety'' properties, where one must verify all possible execution paths of the DUT to ensure that the property is not violated\blue{; and} \red{.The complexity of verifying such properties grows exponentially with the size of the DUT, as there can be an exponential number of paths that need to be checked.}
(ii)~``cover/progress/trace-existing'' properties, where one must verify there exists a path from the initial state of the DUT to a state where the property is satisfied. 
\blue{The relationship between a \textit{cover} property and an \textit{assert} property can be shown as $cover(p) = \neg (assert(\neg p))$, where $p$ represents the expression of a property. Hence, compared to symbolic execution, formal tools such as \JG{} that support \textit{cover} property can (i)~explore hard-to-reach regions based on various coverage metrics rather than execution paths and (ii)~efficiently verify larger DUTs since the tools only need to find the existence of one path that satisfies the property.
}

\red{Compared to {\it assert} properties, {\it cover} properties can be efficiently verified for much larger DUTs since the tools only need to find the existence of a path that satisfies the property. 
Furthermore, commercial tools like Cadence \JG{} already support verification of {\it cover} properties for hardware~\cite{jaspergold}. 
Hence, we propose to use {\it cover} properties for \ourtool{}.}

On \red{verifying}\blue{proving} a \blue{\textit{cover}} property, formal tools will return one of three results: (i)~\textit{unreachable}, (ii)~\textit{reachable}, or (iii)~\textit{undetermined}~\cite{punnoose2014survey}.
A property is \textit{unreachable} when there is no \blue{execution} path from an initial state of the DUT to a state satisfying the property. 
A \textit{reachable} property will have at least one such path. Formal tools usually output such a path consisting of Boolean assignments for the inputs of the DUT for each clock cycle to satisfy the property. \ourtool{} uses such assignments to generate the test cases as seeds for the fuzzer (see Section~\ref{sec:basic_framework}). 
A property is \textit{undetermined} if the formal tool cannot find a path within a given time limit. 
We account for all these properties while building \ourtool{}.

Several commercial formal tools verify \blue{hardware} DUTs, such as Siemens \questa{}~\cite{questa}, Synopsys \vcformal{}~\cite{vcformal}, and Cadence \JG{}. They operate on hardware designs represented in different hardware description languages~(HDLs). 
For \ourtool{}, we \blue{currently} use the \red{widely-used }\JG{}\red{ tool}, which is well-known for its performance and features supported~\cite{punnoose2014survey} and is also used for the verification of RISC-V processors~\cite{RISCVJGUsage}.

Despite their performance, these tools cannot formally verify complete processor designs \blue{because the size of the designs and/or complexity are often too big}. 
For instance, \JG{} took around eight days to verify $94.51\%$ of the branch coverage points in \cva{} processor~\cite{cva6} (see Section~\ref{sec:formal_limitations}), motivating the need for techniques such as hardware fuzzing.

\blue{\JG{} includes SAT/BDD-based formal engines with variations of these algorithms to prove properties~\cite{jgUserManual}. Therefore, \ourtool{} could use any other SAT/BDD-based proof engines that support the generation of Boolean assignments for SVA properties to generate test cases.}

\subsection{Hardware Fuzzing}

Most hardware fuzzers consist of a \textit{seed corpus}, \textit{mutator}, and \textit{vulnerability detector}~\cite{hur2021difuzzrtl, kandethehuzz,fuzzhwlikesw}.
The \textit{seed corpus} is an initial set of input test cases called \textbf{seeds}~\cite{hur2021difuzzrtl}. 
These input \textbf{test cases} are the inputs required to simulate the DUT. 
The \textit{seed corpus} is either manually crafted or generated randomly~\cite{muduli2020hyperfuzzing}. The fuzzer simulates the DUT with these test cases, collects \textbf{coverage}, and \textbf{mutates} all ``interesting'' test cases (i.e., test cases that \red{increase}\blue{achieve} coverage) using its \textit{mutator} to generate new test cases~\cite{rfuzz}. 
The \textit{vulnerability detector} reports any vulnerabilities detected during the simulation. 
The fuzzer simulates these new test cases and repeats the cycle until it achieves the desired coverage. Next, we explain the various components and tasks performed by hardware fuzzers.

\noindent\textbf{DUT} is a hardware design written in HDLs like \verilog{} and \systemverilog{}~\cite{muduli2020hyperfuzzing,kandethehuzz,fuzzhwlikesw} or hardware construction languages~(HCLs) like Chisel~\cite{rfuzz,hur2021difuzzrtl,ragab_bugsbunny_2022,canakci2021directfuzz}. Hardware fuzzers use simulation tools like \verilator{}~\cite{verilator}, Synopsys \vcs{}~\cite{synopsys_home}, and Siemens \modelsim{}~\cite{modelsim} to simulate these DUTs. 

\noindent\textbf{Test cases} of generic hardware fuzzers include data for each input signal of the DUT for each clock cycle~\cite{rfuzz,fuzzhwlikesw,muduli2020hyperfuzzing}. In contrast, fuzzers designed specifically to fuzz processors generate binary executable files as test cases~\cite{hur2021difuzzrtl,kandethehuzz}.

\noindent\textbf{Coverage} measures the number of various types of hardware behaviors, such as toggling the select signals of muxes (mux-toggle coverage~\cite{rfuzz}) and setting registers that drive the selected signals of muxes to different values (control-register coverage~\cite{hur2021difuzzrtl}) during the simulation\red{ of the DUT}. 
\textbf{Coverage points} are assigned to each of these behaviors.
For example, branch coverage indicates whether the different paths of a branch statement are covered or not. Whenever a design enters one of the branch paths, its corresponding coverage point is considered \textbf{covered}; otherwise, it remains \textbf{uncovered}. 

\red{Some fuzzers~\cite{kandethehuzz,fuzzhwlikesw}  use one or many of the code coverage metrics, such as state transitions in the finite-state machine~(FSM), expressions in combinational logic, conditions in branch statements, and bit-toggle events in flip-flops.}
The DUT is instrumented to generate coverage during the simulations~\cite{rfuzz}. 
Thus, covering all the coverage points in the DUT is essential to verify all the hardware behaviors.
Hardware fuzzers use coverage as \textbf{feedback} to determine the interesting test cases~\cite{rfuzz, hur2021difuzzrtl,kandethehuzz,fuzzhwlikesw}.


\noindent\textbf{Mutations} are data manipulation operations, such as bit-flip, byte-flip, clone, and swap, inspired by software fuzzers like the AFL fuzzer~\cite{citeafl}~\cite{rfuzz, kandethehuzz}.

\noindent\textbf{Vulnerability detection} in hardware fuzzers involves either differential testing or assertion \red{generation}\blue{checking}. In differential testing, the outputs of the DUT and a golden reference model~(GRM), when tested with the same test case, are compared to detect vulnerabilities~\cite{kandethehuzz,hur2021difuzzrtl}. In assertion checking, we insert the conditions to trigger the vulnerabilities or assertion properties into the DUT based on its specification and use the violations of these assertions during the simulation to detect vulnerabilities~\cite{muduli2020hyperfuzzing,fuzzhwlikesw}. Note that, unlike software, hardware does not have events like crashes, memory leaks, \blue{and }buffer overflows\red{, and return codes} to use for vulnerability detection~\cite{fuzzhwlikesw}. 

%% file: openarch/motivation.tex
\section{Motivation}\label{sec:motivation}
In this section, we highlight the limitations of existing formal and fuzzing techniques to motivate the need for hybrid hardware fuzzers. 
To this end, we use a popular, open-sourced RISC-V~\cite{riscv_home} based processor, \cva{}~\cite{cva6}, as a case study. However, we perform extensive evaluation of \ourtool{} on all modules of five different processors (see Section~\ref{sec:exp_results}).

\subsection{Case Study on \texttt{\textbf{CVA6}} Processor}
 
Consider the Listing~\ref{listing:case_cva6_interrupt}, which shows the trigger condition of three interrupts in the interrupt handler of \cva{}. 
Verifying the correctness of the interrupt handler is critical, as a vulnerable interrupt handler can be exploited for information leakage~\cite{de2014secure}.
To trigger each type of interrupt, the corresponding bits of two control \red{state}\blue{and status} registers~(CSRs): \texttt{mie} and \texttt{mip}, need to be enabled (i.e., set to \texttt{1'b1}). 
Thus, the test case should simultaneously consists of instructions that set the bits of both \texttt{mie} and \texttt{mip} registers. 
This condition is covered by the {\it branch coverage} metric, which checks \red{whether}\blue{if} both directions of the branch (in this case, the {\it if} statement) are taken~\cite{mockus2009test}.

\subsection{Limitations of Existing Hardware Fuzzers} \label{sec:fuzzer_limitations}
Hardware fuzzers iteratively perform seed generation and mutation to improve coverage~\cite{rfuzz, kandethehuzz, hur2021difuzzrtl,fuzzhwlikesw,ragab_bugsbunny_2022, canakci2021directfuzz}. 
However, fuzzers still require an exponentially large amount of time to cover some coverage points---whose test cases are hard to generate due to the specific conditions required to trigger them---leaving multiple design spaces unexplored and vulnerabilities undetected.

\noindent\textbf{Case Study on \texttt{\textbf{CVA6}}'s interrupt controller.} We fuzzed the \cva{} processor with the most recent processor fuzzer, \thehuzz{}, 
for 72 hours, generating more than $200K$ test cases\footnote{For this paper, a \textit{test case} refers to a binary executable.}. 
Unfortunately, \thehuzz{} did not cover any of the branch coverage points of all the three interrupts~\cite{kandethehuzz}.
We performed further analysis to understand this limitation. 

Consider the \red{\texttt{S\_TIMER}}\blue{\texttt{S\_EXT}} interrupt in Line \red{5}\blue{9} in Listing~\ref{listing:case_cva6_interrupt}. 
Triggering this interrupt requires the \texttt{S\_\red{TIMER\_INTERRUPT}\blue{EXT\_INTERRUPT}} bit of both the CSRs, \texttt{mie} and \texttt{mip}, to be enabled. According to the RISC-V instruction set architecture (ISA) emulator, \spike{}~\cite{spike}, there are only four instructions ({\tt CSRRW}, {\tt CSRRWI}, {\tt CSRRS}, and {\tt CSRRSI}) out of the total $1146$ RISC-V instructions \blue{can modify the values of CSRs}, and there are $229$ CSRs in total. The probability of generating an instruction that sets the \texttt{mie} register is $ 4/(1146 \times 229) = 1.524 \times 10^{-5}$. 
As we also need the bit of \texttt{mip} CSR to be set\red{ simultaneously}, the combined probability of generating such a test case is only $(1.524 \times 10^{-5}) ^ 2 = 2.323 \times 10^{-10}$. 
Though this is the probability of randomly generating a test case, it sheds light on why existing hardware fuzzers ~\cite{hur2021difuzzrtl, kandethehuzz, canakci2021directfuzz, ragab_bugsbunny_2022}---which use the coverage feedback and mutation techniques---could not cover this coverage point \red{even }after fuzzing for 72 hours. 
This demonstrates that \textit{existing hardware fuzzers are insufficient to explore hard-to-reach design spaces, leaving vulnerabilities undetected.} 

\begin{lstlisting}[language=Verilog, label = {listing:case_cva6_interrupt}, caption={\red{Verilog code snippet of the }Interrupt handler in the \cva{} processor~\cite{cva6}.},style=prettyverilog,float,belowskip=-15pt,aboveskip=0pt,firstnumber=1,linewidth=\linewidth]
localparam SupervisorIrq = 1;
localparam logic [63:0] S_EXT_INTERRUPT = 64'd9;
. . .
// Interrupt handler
if (mie[S_TIMER_INTERRUPT] && mip[S_TIMER_INTERRUPT])
    interrupt_cause = S_TIMER_INTERRUPT; // Supervisor Timer
if (mie[S_SW_INTERRUPT] && mip[S_SW_INTERRUPT])
    interrupt_cause = S_SW_INTERRUPT;    // Supervisor Software
if (mie[S_EXT_INTERRUPT] && (mip[S_EXT_INTERRUPT] | irq[SupervisorIrq]))
    interrupt_cause = S_EXT_INTERRUPT;   // Supervisor External
\end{lstlisting}

\subsection{Limitations of Formal Verification}\label{sec:formal_limitations}
Theoretically, formal tools can use \textit{cover} properties to \red{verify}\blue{prove the reachability of} all the coverage points in a DUT~\cite{IEEEstd}, achieving 100\% \blue{of the reachable} coverage.
However, this requires writing \blue{and proving} \textit{cover} properties for all the coverage points, an error-prone and time-consuming task (as it requires design knowledge and manual effort), \red{and verifying these properties with formal tools, a resource-intense task, }especially in large and complex hardware designs like processors with thousands of coverage points~\cite{dessouky2019hardfails}. \blue{Moreover, since the reachability of a \textit{cover} property only requires the existence of one path, formal tools, like \JG{}, do not guarantee to find all the paths with vulnerabilities.} 

\noindent\textbf{Case Study on \texttt{\textbf{CVA6}}'s interrupt controller.}
The \cva{} processor has $9.53\times10^3$ branch coverage points. 
First, there exists no tool that can convert these branch coverage points into \textit{cover} properties that tools like \JG{}~\cite{jaspergold} can \red{verify}\blue{prove}. Thus, one has to craft these properties manually. We developed and used a \textit{property \blue{generator}\red{converter}} that can  automatically derive \textit{cover} properties from  branch coverage points. 

\blue{Second, but more importantly, we evaluate the time taken by \JG{} to prove all these properties and the coverage it achieves.
Since the corresponding Boolean assignments generated by \JG{} cannot be directly used as test cases for fuzzing a processor, we developed a \textit{test case converter} that can automatically convert these Boolean assignments into the binary executable format. 
We then simulate these test cases and collect the branch coverage achieved. The time consumption of each point is the summation of formal verification and simulation. To prevent \JG{} from spending too much time on the property of one point, we limit the maximal time on each property~(see Section~\ref{cc:exp_setup}). \JG{} took eight days to verify $94.51\%$ of the branch points in the \cva{} processor as shown in Figure~\ref{fig:whycombine}.}
\red{Second but more importantly, \JG{} took eight days to verify $94.51\%$ of the branch points in the \cva{} processor.}
This shows that \textit{using formal tools requires extensive manual labor and has tremendous runtime in verifying all coverage points in the DUT.}

\noindent\blue{\textbf{Case Study on vulnerability detection.} Theoretically, formal tools alone can achieve 100\% coverage by proving the \textit{cover} properties of all the coverage points. But, there is still scope for vulnerabilities. For example, consider the vulnerability~\ref{v3} found by \ourtool{} where the \cva{} processor returns X-values when accessing unallocated CSRs.
Listing~\ref{listing:case_why_formalfuzz} is a 
code segment from the \cva{} processor that accesses data from the hardware performance counters~(HPCs) using the address of the CSRs~(\texttt{csr\_address}). The HPCs are used for anomaly and malicious behavior detection~\cite{wang2013numchecker, krishnamurthy2019anomaly}, hence verifying their security is essential. Triggering the vulnerability~\ref{v3} requires a test case to access the data of the counters among \texttt{MHPM\_COUNTER\_17} to \texttt{MHPM\_COUNTER\_31}. 
However, \JG{} will not always find a path to access the target counters. This is because all counters share the same point to reduce the instrumentation overhead from the branch coverage metric. A \textit{cover} property \blue{does not} explore all paths under the point, which shows that \textit{using formal tools with standard \textit{cover} properties is insufficient to detect all vulnerabilities.}} 

\begin{lstlisting}[language=Verilog, label = {listing:case_why_formalfuzz}, caption={The Verilog code of CSR reading in  \cva{}.},style=prettyverilog,float,belowskip=-0pt,aboveskip=-5pt,firstnumber=1]
if (csr_read) begin
  unique case (csr_address)
    // Counters and Timers
    ML1_ICACHE_MISS, ML1_DCACHE_MISS, MITLB_MISS, 
    MDTLB_MISS, MLOAD, MSTORE, MEXCEPTION, 
    MEXCEPTION_RET, MBRANCH_JUMP, MCALL, MRET,
    MMIS_PREDICT, MSB_FULL, MIF_EMPTY,
    MHPM_COUNTER_17,
    ...
    MHPM_COUNTER_31:           csr_rdata   = perf_data_i; => Point
\end{lstlisting}

\subsection{Advantages of a Hybrid Fuzzer}\label{sec:advantagesOfHyPFuzz}
By using both \red{tools}\blue{techniques} in tandem, hybrid fuzzers overcome the limitations of using fuzzing and formal \red{tools}\blue{techniques} alone. The fuzzer quickly explores the DUT through the mutation of effective test cases. The formal tool verifies points that \blue{the} fuzzer is struggling to cover and provides the corresponding test cases as seeds to the fuzzer. \blue{The fuzzer then mutates these test cases to explore the target design further.}

\noindent\textbf{Case Study on \texttt{\textbf{CVA6}}'s interrupt controller.}
Since the fuzzer can cover none of the three coverage points in the interrupt handler (see Section~\ref{sec:fuzzer_limitations}), the hybrid fuzzer uses a formal tool \red{to cover these points}\blue{for assistance}. 
\red{Using \JG{}, we verified the coverage point of \texttt{S\_TIMER} interrupt in only 20 seconds.}
\blue{Of these three branch coverage points, consider the point of \texttt{S\_EXT\_INTERRUPT} at line 9 in Listing~\ref{listing:case_cva6_interrupt} as an example. 
We first use the conditions for covering this point~(see Appendix~\ref{apd:cov_metric}) into an SVA \textit{cover} property:
\texttt{cover property} $(mie[S\_EXT\_INTERRUPT]$ $~\&\&$ $~mip[S\_EXT\_INTERRUPT]$ $~||$ $~irq[SupervisorIrq])$. \JG{} then takes only 20 seconds to find a path to this property and dump Boolean assignments as shown in Figure~\ref{fig:jg_boolean_assign}.}

\blue{However, the fuzzer cannot directly use these Boolean assignments because the processors require binary executables as test cases. Hence, we identify instruction-related signals~(e.g., the input instruction port of the decoder) from the Boolean assignments and parse their values beginning from the initial state to the state that satisfies the property.
Listing~\ref{listing:case_instr} shows the extracted sequence of instructions.}

\begin{lstlisting}[language=Verilog, label = {listing:case_instr}, caption={Instructions covering the point of the \texttt{S\_EXT\_INTERRUPT}.},style=prettyverilog,float,belowskip=-5pt,aboveskip=0pt,firstnumber=1]
    ORI X6, X3, h'204;    // update reg X6 with s_ext value
    CSRRS X0, mie, X6; // write the value of X6 to mie
    CSRRS X0, mip, X6; // write the value of X6 to mip
\end{lstlisting}

\blue{
Then, the sequence of instructions is converted into a valid executable file. Such files consist of three instruction sequences: \texttt{INIT} instructions that initialize registers and memory of the processor, \texttt{TEST} instructions that contain the testing instruction sequence, and \texttt{EXIT} instructions that handle normal and abnormal~(e.g., exception) termination of simulation, as shown in Figure~\ref{fig:case_test_case}.
To generate a valid executable file, we create an executable file template with \texttt{NOP} instructions as \texttt{TEST} instructions, compare and identify the initial memory address of the \texttt{TEST} instruction section from the disassembly file, and replaces these \texttt{NOP} instructions with the instruction sequence extracted from the Boolean assignments.
}




The hybrid fuzzer can now use this test case as a seed, mutate it\blue{,} and generate new test cases that cover the remaining two coverage points in the interrupt handler. 
For example, a commonly-used mutation technique in existing hardware fuzzers~\cite{hur2021difuzzrtl, kandethehuzz, canakci2021directfuzz, ragab_bugsbunny_2022}, \textit{random-8}, overwrites a random byte with a random value in the instruction~\cite{kandethehuzz, rfuzz}.
This mutation can easily toggle the other bits in \texttt{mie} and \texttt{mip} CSRs, covering the other coverage points of the interrupt handler.  
Therefore, a hybrid fuzzer can \textit{use test cases from formal tools to overcome the limitation of fuzzers, thereby increasing the coverage}.    


\begin{figure}
    \centering
    \includegraphics[width=0.85\linewidth]{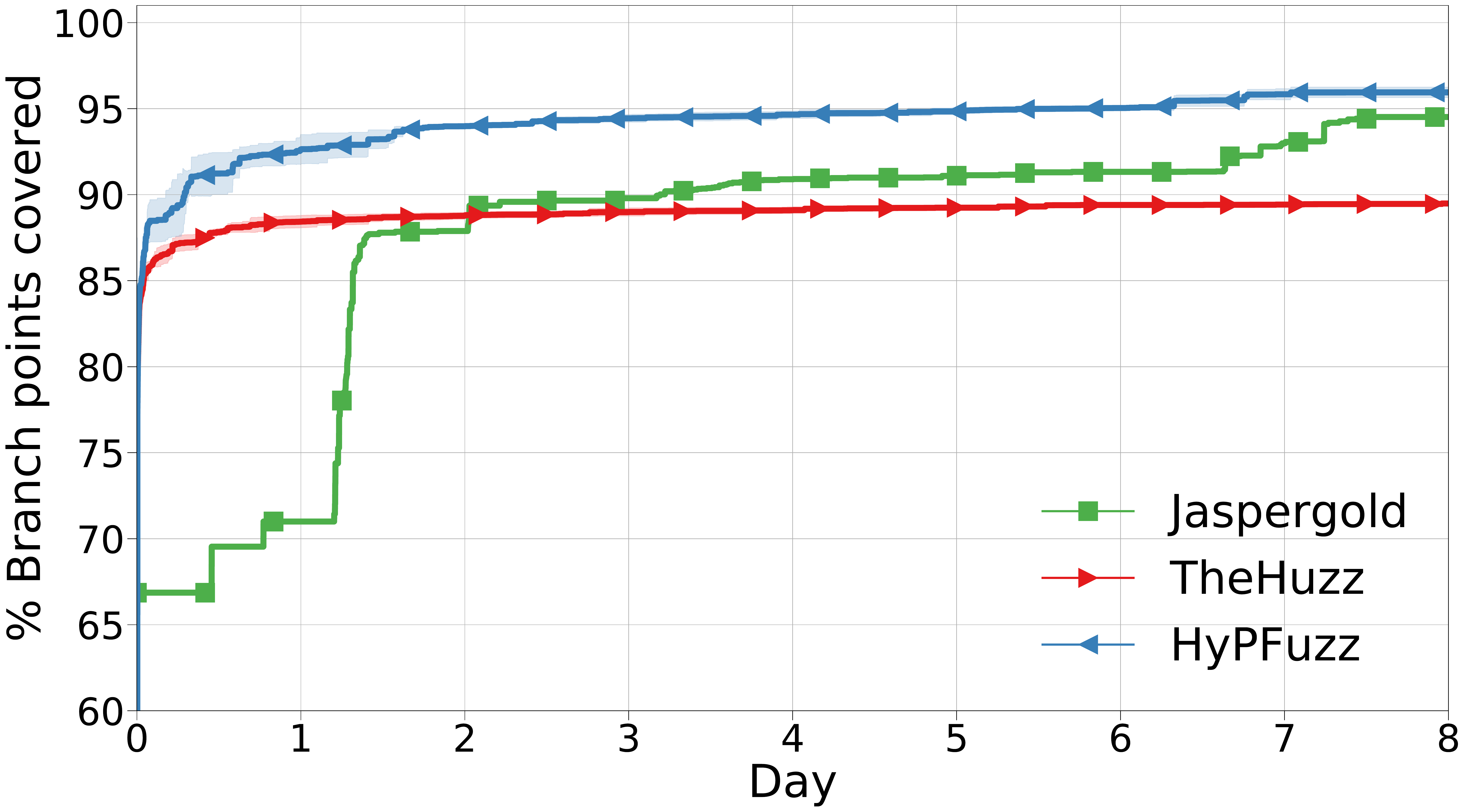}
    \caption{Eight day coverage results of \thehuzz{}~\cite{kandethehuzz}, \JG{}~\cite{jaspergold}, and \ourtool{} for the \cva{} processor~\cite{cva6}. }
    \label{fig:whycombine}
\end{figure}

We built a \textit{property generator}\red{(see Section~\ref{sec:integrateFF})} to automatically generate the \textit{cover} properties of \red{branch}coverage points \blue{and a \textit{test case converter} (see Section~\ref{sec:integrateFF}) to automatically convert Boolean assignments of reachable properties into test cases.}
\red{Since verifying these properties has a huge run-time overhead (see Section~\ref{sec:formal_limitations}), the hybrid fuzzer uses the fuzzer for most of the design exploration and the formal tool only to verify the points not covered by the fuzzer.} 

\noindent \textbf{Case Study on \texttt{\textbf{CVA6}}.}
Figure~\ref{fig:whycombine} shows the branch coverage achieved by the most recent processor fuzzer\red{--\thehuzz{}}\blue{, \thehuzz{};} the formal tool\red{--\JG{}}\blue{, \JG{};} and our hybrid fuzzer, \ourtool{}. 
 \JG{} continues to \red{increase}\blue{achieve} coverage but is slow due to its high run-time to cover each coverage point; it reaches a coverage of $94.51\%$ after running for eight days\red{ (not shown in the figure)}. 
On the other hand, even though \thehuzz{} initially achieves faster coverage than \JG{}, it fails to \red{increase}\blue{achieve} the coverage beyond $88\%$, even after running for $72$ hours, as all the remaining coverage points are hard-to-reach. 
In contrast, \ourtool{} achieves $94.78\%$ coverage ($6.1\%$ more coverage compared to \thehuzz{}) in $72$ hours, and it achieves the $94.51\%$ coverage achieved by \JG{} in $50.71$ hours ($3.71 \times$ faster than \JG{}). 
Therefore, a hybrid fuzzer can \textit{use the fuzzer to explore the DUT and overcome the manual and run-time overhead limitations of formal tools,  achieving coverage faster}.

%% file: openarch/method.tex
\section{Hybrid Hardware Fuzzing}

In this section, we first elaborate on the challenges of building a hybrid fuzzer and how we address them in \ourtool{}. 

\subsection{Challenges}

\begin{chal}[\label{c1}] \textbf{Scheduling:}
The speed of the fuzzer varies over time depending on the design-under-test (DUT) and the type of fuzzer used. Similarly, the speed of the formal tool varies from one coverage point to another based on the DUT, the type of formal tool, and the computational resources\red{ available}. 
\red{Consider a trivial scheduling strategy: alternating between the formal tool and the fuzzer after a fixed amount of time. This strategy will not be optimal because the fuzzer may fail to increase the coverage but continue to run, or one may switch to the formal tool even when the fuzzer is quickly increasing the coverage.}  
        Thus, challenge~\ref{c1} is to \textit{build \red{an optimal}\blue{a dynamic} scheduler between the formal tool and the fuzzer that minimizes the under- and over-utilization of the capabilities of the formal tool and the fuzzer.}
\end{chal}

\begin{chal}[\label{c2}] 
\textbf{Selection of coverage points:}
Since the coverage point targeted by the formal tool
determines the seed of the fuzzer, it also impacts the successive points covered by the fuzzer as it mutates this seed. 
Thus, the hybrid fuzzer should select the uncovered points that maximize the number of coverage points the fuzzer can uncover, thereby increasing the speed of the fuzzer. 
\red{For instance, one may select an uncovered point in a module with more uncovered points than a module with fewer or no uncovered points.} 
However, the current set of uncovered points depends on the DUT and also what points have been covered by the fuzzer in the past.
Thus, challenge~\ref{c2} is to \textit{build a  point selector that maximizes the rate of coverage despite the uneven distribution of uncovered points in the DUT, which also changes with time.}
\end{chal}

\begin{chal}[\label{c3}] 
\textbf{Seamless integration.}
A hybrid fuzzer should seamlessly integrate the formal tool and the fuzzer to be faster and \blue{easier} to use.
However, the inputs and outputs of the fuzzer and formal tool are incompatible. 
The fuzzer uses test cases as input, while the formal tool generates Boolean assignments for the inputs of the DUT for each clock cycle as the output. Also, the formal tool needs \textit{cover} properties as input while the fuzzer outputs coverage of each point in the DUT. 
Hence, it is not straightforward to combine a formal tool and a fuzzer. 
\red{Furthermore, no existing technique can automatically convert the target coverage point  to \textit{cover} properties or Boolean assignments of individual inputs to test cases. }
Hence, challenge~\ref{c3} is to \textit{seamlessly integrate formal and fuzzing tools to build an automated flow for the hybrid fuzzer.}
\red{The software fuzzing community has faced similar challenges~\cite{stephens2016driller}, but due to the fundamental differences between hardware and software~\cite{fuzzhwlikesw}, such solutions cannot be adopted to \ourtool{}.}

\end{chal}

We address challenges \ref{c1} and \ref{c2} by building a dynamic \textit{scheduler} and an \textit{uncovered point selector}, respectively.
We solve \ref{c3} by building a \textit{property generator} and a \textit{test case converter}, which facilitate seamless integration of the fuzzer with the formal tool.

\begin{figure}
    \centering
    \includegraphics[trim=18 15 18 18,clip,width=0.85\linewidth]{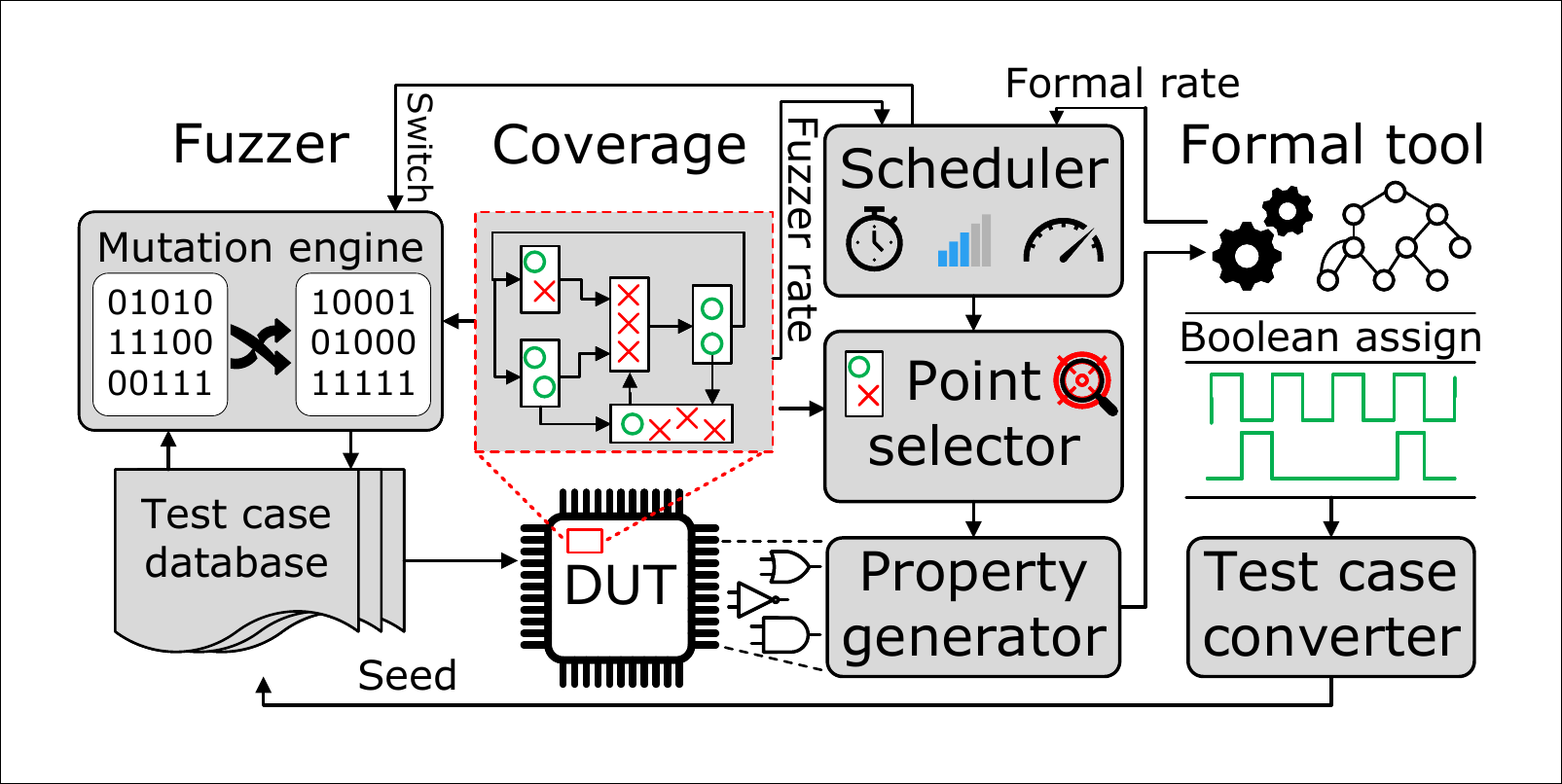}
    \caption{Framework of \ourtool{}. The circle and cross represent covered and uncovered points, respectively.}
    \label{fig:frame}
\end{figure}

\subsection{Framework of \ourtool{}}\label{sec:basic_framework}

\ourtool{} consists of the \textit{scheduler}, \textit{point selector}, \textit{property generator}, and \textit{test case converter}, apart from the fuzzer and the formal tool, as shown in Figure~\ref{fig:frame}. \ourtool{} starts by invoking the \textit{point selector} that \red{optimally}\blue{heuristically} selects the uncovered point that the formal tool should verify (see Section~\ref{sec:selector}).
Then, the \textit{property generator} generates the \textit{cover} property for that point (see Section~\ref{sec:integrateFF}). The formal tool \red{verifies}\blue{proves} this property and generates the Boolean assignments for each input of the DUT for every clock cycle required to trigger that uncovered point.
The \textit{test case converter} converts the Boolean assignments of \red{input}\blue{instruction} signals into a test case, which the fuzzer uses as a seed (see Section~\ref{sec:integrateFF}). 
The fuzzer simulates the DUT with this seed to reach the coverage point.
The fuzzer also mutates this seed to cover the neighborhood points. 
It runs until the \textit{scheduler} stops it. Then, the \textit{point selector} selects the next \red{optimal}uncovered point. 
\ourtool{} repeats this process until it achieves the target coverage or hits a timeout.

\subsection{\red{Optimal }Scheduling of Fuzzer and Formal Tool}\label{sec:scheduler}
Our \textit{scheduler} uses a dynamic scheduling strategy that switches \ourtool{} from fuzzer to formal tool  when the rate of coverage increment of the fuzzer ($r_{fuzz}$) is less than that of the formal tool ($r_{fml}$).
\blue{The rates reflect their capability to explore the design spaces.}
On the other hand, the formal tool is \red{run}\blue{running} until it generates the Boolean assignments to reach the \red{coverage}\blue{uncovered} point.
Next, we formulate the rate of coverage increment of the fuzzer and formal tool, which helps one to schedule them \red{optimally}\blue{dynamically}. 

\noindent\textbf{Formal tool's coverage increment rate ($r_{fml}$)} is the ratio of the number of coverage points verified by the formal tool so far. 
\blue{Since a formal tool will process the points that are hard to be covered by the fuzzer, we can calculate the optimal $r_{fml}$ using $\frac{n}{t_p}$, where $n$ represents the uncovered points selected when switching to the formal tool, and $t_p$ represents the time spent by the formal tool on proving the corresponding properties. However, since we need to calculate $r_{fml}$ before switching to the formal tool, \ourtool{} needs to predict the uncovered points selected and the time taken by the formal tool on the corresponding properties. Unfortunately, both predictions are difficult due to the randomness of fuzzing and the low prediction accuracy of the time cost of a formal tool (the accuracy of the state-of-the-art machine learning strategy is $68\%$, given a property~\cite{el2016estimation}). Therefore, we first calculate the average time $t_{ave}$ spent by a formal tool on properties. We then use $t_{ave}$ to estimate the $r_{fml}$ as $\frac{1}{t_{ave}}$ to reflect how the formal tool will explore the hard-to-reach spaces of the fuzzer in a design.}

\blue{However, it is infeasible to calculate $t_{ave}$ by proving all uncovered points in the hard-to-reach region. Therefore, we estimate $r_{fml}$ using the moving average, which is widely applied to estimate the underlying trend~\cite{hyndman2011moving}, as}
\red{Let $\mathcal{C}$ be the set of coverage points verified by the formal tool, and let $t_{fml}(c)$ denote the run-time of the formal tool to generate a test case for the coverage point $c$. Thus,}

\begin{equation}\label{eq:r_ideal_formal}
    r_{fml} = \frac{\left | \mathcal{C} \right |}{\displaystyle\sum_{c \in \mathcal{C}} t_{fml}(c)}, \tag{1}
\end{equation}
\blue{where $\mathcal{C}$ is the set of coverage points verified by the formal tool, and $t_{fml}(c)$ denotes the run-time of the formal tool to generate a test case for the coverage point $c$.}

\noindent\textbf{Fuzzer's coverage increment rate ($r_{fuzz}$)} is the ratio of the number of \blue{new} coverage points covered by the fuzzer in a rolling window over the run-time of the fuzzer. 
\red{$r_{fuzz}$ decreases as more points are uncovered because a fuzzer will spend more time exploring hard-to-reach design spaces.}\blue{To prevent under-/over-utilization of a fuzzer, $r_{fuzz}$ reflects how well the fuzzer recently explored the design spaces. Fuzzers usually achieves faster coverage initially and then slow down due to the hard-to-reach spaces in the design~\cite{rfuzz,kandethehuzz,hur2021difuzzrtl}. Therefore, if we calculate $r_{fuzz}$ including coverage increment at the beginning, $r_{fuzz}$ will become unnecessarily high and cause over-utilization of the fuzzer, delaying the switching process. However, if we calculate $r_{fuzz}$ using the coverage achieved by the most recent test case, the test case may not achieve new coverage, whereas the upcoming test cases can due to the randomness of the fuzzing process. This causes under-utilization of the fuzzer and hence cannot fuzz around the seed from the formal tool entirely.} \red{Hence}\blue{Therefore}, a rolling window is used in this case to compute the instantaneous rate of the fuzzer.
Let $K$ denote the set of all the test cases generated by the fuzzer, where $k_i$ denotes the $i^{th}$ test case generated. Then, the set of test cases in the window $w$ is $K_w = \{ k_i \in K \mid |K|-w < i \leq |K|\}$. 
Let $n(k_i)$ and $t_{fuzz}(k_i)$ denote the number of new coverage 
points covered and the run-time of the fuzzer for each test case, respectively. Thus,

\begin{equation}\label{eq:r_fuzzing}
 r_{fuzz}(w) =  \frac{ 
\displaystyle\sum_{k_i \in K_w}  n(k_i)
}
{
\displaystyle\sum_{k_i \in K_w} t_{fuzz}(k_i) 
} \tag{2}
\end{equation}

The \textit{scheduler} computes $r_{fuzz}$ and $r_{fml}$ in real-time (see Section~\ref{cc:exp_setup}), runs the fuzzer as long as it can cover more points than the formal tool, and then switches to the formal tool when $r_{fuzz} < r_{fml}$. 

\begin{figure*}[tb!]
    \centering
    \includegraphics[trim=18 18 20 19,clip,width=1.7\columnwidth]{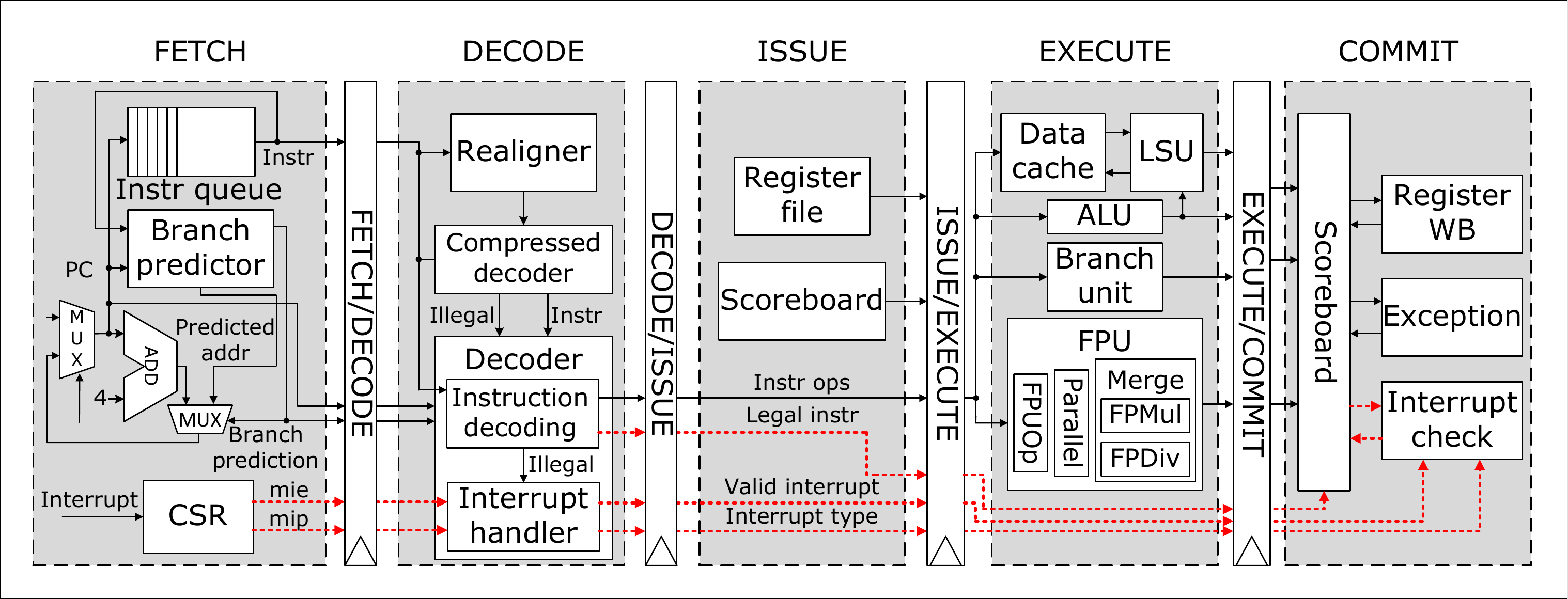}
    \caption{Simplified pipeline of the \cva{} processor~\cite{cva6}. 
    The connectivity in dotted lines is an example to show how a module drives various other modules in the design.}
    \label{fig:processorPipeline}
\end{figure*}

\subsection{\red{Optimal }Selection of Uncovered Points}\label{sec:selector}
The hardware design and verification process is modular~\cite{modularDesign,canakci2021directfuzz}.
Therefore, to be compatible with the existing hardware design verification flow, we develop strategies for the \red{optimal}selection of uncovered points at the module level. 

We run existing processor fuzzers, including \difuzz{}~\cite{hur2021difuzzrtl} and \thehuzz{}, on up to five different processors and analyze their effectiveness at the module level. 
Based on this study, we make the following observations:

\begin{obs}\label{distObs}.~The farther the module is from the inputs of the DUT, the harder it is for the fuzzer to trigger the module's components accurately. Thus, the coverage of the fuzzer is\red{ directly }proportional to the {\it distance} of the module from the input. \end{obs}
\begin{obs}\label{noUncObs}. The more the number of uncovered points in the module, the more coverage points the fuzzer can \red{uncover}\blue{cover} if the fuzzer's seed activates the module. \end{obs}
\begin{obs}\label{depObs}. The more DUT logic a module drives, the higher the probability of the fuzzer \red{uncovering}\blue{covering} new points. 
\red{To this end, we use the fanout cone-of-influence (COI), a traditional metric used in the hardware verification process, to indicate the amount of DUT logic a module drives~\cite{el2016estimation}. The fanout COI of a module is the sum of the number of input signals of other modules driven by the target module.}

\end{obs}

Based on these observations, we have developed three strategies\red{to select points}: (i)~\bottop{}, (ii)~\maxuncov{}, and (iii)~\moddep{}.
\blue{The strategies include deterministic and non-deterministic operations, which will first select a module and then randomly select an uncovered point inside.}
We also use \randsel{}, a naive selection strategy that randomly selects the \red{modules}\blue{uncovered points}.
We evaluate these strategies on a comprehensive set of five real-world, open-source processors covering one of the first and most widely used OpenRISC ISA~\cite{openrisc_home} and RISC-V ISA~\cite{riscv_home}, respectively, and select the best strategy for our \textit{point selector} based on empirical results (see Section~\ref{sec:covAnalysis}).
We now explain these three strategies using \cva{} as \red{a running example}\blue{an example}.

\subsubsection{\bottop{} Strategy}\label{sec:botTop}
To ensure the \red{optimal}\blue{efficient} usage of the formal tool and fuzzer, the formal tool should target the hard-to-reach coverage points, and the fuzzer should target the remaining points.
Based on Observation~\ref{distObs}, \bottop{} selects modules {\it deep} in the DUT, i.e., their distance from the DUT's input. 
We define this distance as the number of modules between the input of the DUT and the target module. 
For example, consider  the simplified pipeline of the \cva{} processor shown in Figure~\ref{fig:processorPipeline}. In this pipeline, the distance of {\tt FETCH}, {\tt DECODE}, and {\tt EXECUTE} is one, and both {\tt FPMul} and {\tt FPDiv} have a distance of four.  
Thus, the \bottop{} strategy assigns the highest priority to {\tt FPMul} and {\tt FPDiv} over other modules in the \cva{} processor. 

\red{\noindent\textbf{Limitation.} 
By selecting  a module that is {\it deep}, \bottop{} ensures that the formal tool targets a coverage point that is {\it deep} in the DUT. However, the \bottop{} strategy does not guarantee that the fuzzer is used optimally because  a {\it deeper} module may not have any other uncovered point apart from the one targeted by the formal tool, or it may have few coverage points, and thus, under-utilizing the fuzzer. In fact, for many processors, the deepest modules are flip-flops (FFs), buffers, or simple combinational components with few coverage points.} 


\red{For example, in the \cva{} processor, \bottop{} assigns highest priority to \texttt{iteration\_div\_sqrt} module, which has only two branch coverage points.
This is very low compared to the average number of branch coverage points across all modules in the \cva{} processor, which is $29.60$. 
Thus, \bottop{} optimizes the utilization of the formal tool but not the fuzzer. Our next selection strategy, \maxuncov{}, addresses this issue.}

\subsubsection{\maxuncov{} Strategy}\label{sec:maxuncov}
In a hybrid fuzzer, the seed generated using the formal tool will cover a coverage point in the hard-to-reach design space. 
On mutating the seed, the fuzzer will explore the design space in the ``vicinity'' of this covered point. 
Thus, based on Observation~\ref{noUncObs}, having more uncovered points  in the ``vicinity''  will increase the number of coverage points the fuzzer can cover,  thereby accelerating  \ourtool{}.  
Thus, our second strategy, \maxuncov{} prioritizes the module with the maximum number of uncovered points over the rest of DUT's modules, irrespective of its distance from the inputs.
Unlike the \bottop{} strategy, whose module distance is fixed for a given DUT, the number of uncovered points in a given module decreases  as \ourtool{} explores more design space over time. Hence, \maxuncov{} is a dynamic strategy that recomputes the priorities of each module every time the \textit{point selector} is invoked.

For example, in the \cva{} processor, the \texttt{Decoder}  and floating point unit~(\texttt{FPU}) modules have 381 and 108 branch coverage points, respectively. 
In the beginning, all the coverage points in the DUT are uncovered. Hence, the \texttt{Decoder} will have more uncovered points; consequently, \maxuncov{} prioritizes the \texttt{Decoder} over the \texttt{FPU}. However, over time, the number of uncovered points in the \texttt{Decoder} decreases as instructions of all types activate this module. 

\red{\noindent\textbf{Limitation.} \maxuncov{} selects coverage points with a maximum number of uncovered points in the ``vicinity'' for the fuzzer to cover. However, triggering coverage points in one module depends on the other modules, which drive the input signals of this module. For example, in the interrupt handler of the \cva{} processor (see Section~\ref{sec:motivation}), none of the three branch coverage points were covered when fuzzing with \thehuzz{}~\cite{kandethehuzz} because other modules driving the interrupt handler could not send the input signals required to trigger interrupts in the interrupt controller.  Furthermore, \maxuncov{} needs to compute the number of uncovered points in every module each time the formal tool is invoked; this processing time adds overhead. 
To address these issues, we describe our third \textit{point selector} strategy next.} 

\subsubsection{\moddep{} Strategy}\label{sec:moddep}
Triggering a coverage point first requires activating the modules that drive the logic corresponding to this coverage point.  Thus, based on Observation~\ref{depObs}, targeting the coverage points from a module with many other modules in their fanout COI  will allow the fuzzer to uncover more points. 
Hence, our \moddep{} strategy prioritizes modules with  higher COI over other modules, as shown in Algorithm~\ref{algo:moddep}. 
For example, \moddep{} assigns the highest priority to the \texttt{CSR} module in the \cva{} processor because it has the highest fanout due to driving multiple components: interrupt control logic (shown using red dotted lines in Figure~\ref{fig:processorPipeline}), write logic in \texttt{Register file}, and enable logic in \texttt{Data cache}.
 

\red{\noindent\textbf{Limitation.} The coverage points in the modules with high COI may not always be hard-to-reach. For example, the \texttt{Decoder} in the \cva{} processor has a high COI as it drives multiple components in the \texttt{ISSUE} and the \texttt{EXECUTE} stages. However commonly-used instructions, such as arithmetic and branch, can cover many of \red{its} coverage points. Hence, the coverage points selected by the \moddep{} strategy are not necessarily hard to cover using the fuzzer, resulting in over-utilization of the formal tool.} 

\subsection{Integrating Fuzzer and Formal Tool}\label{sec:integrateFF}
Formal tools target the \red{verification of }\blue{DUT's} properties, but fuzzers target the DUT's coverage points. 
Due to this incompatibility, we cannot directly send the target coverage point to the formal tool to verify or take the Boolean assignments of a \red{verified}\blue{proved} property as an executable test case for the fuzzer, creating challenge~\ref{c3}. \red{To enable seamless integration of}\blue{To seamlessly integrate} the fuzzer and formal tool, we develop a \textit{property generator} and a \textit{test case converter}. 

\noindent\textbf{\textit{Property generator}}\label{sec:propertyGenerator}
generates the \textit{cover} property for the selected uncovered point. It parses the DUT's logic and identifies the conditions for covering the point. Then, it converts the conditions into a \textit{cover} property and loads it to the formal tool along with the DUT. For example, given an uncovered point of branch coverage, the \textit{property generator} analyzes the dependencies of the branch statement. It then identifies the conditions to cover that point, and the logical conjunction of them will form the expression of the corresponding \textit{cover} property. 
\blue{The \textit{property generator} is compatible to uncovered points of other coverage metrics, as shown in Appendix~\ref{apd:cov_metric}.}
\red{\ourtool{} can also apply the \textit{property generator} to uncovered points of other coverage metrics, as shown in Appendix~\ref{apd:cov_metric}.} 

\noindent\textbf{\textit{Test case converter}}\label{sec:testcaseConverter}
As mentioned earlier, 
the formal tool generates the Boolean assignments of the input signals for each clock cycle to cover a target coverage point. 
However, the fuzzer has a different input format. 
For example, processor fuzzers, such as \thehuzz{} and \difuzz{}~\cite{hur2021difuzzrtl}, use sequences of instructions as test cases to fuzz~\cite{kandethehuzz,hur2021difuzzrtl,rfuzz}.
The \textit{test case converter} will use the ISA of the DUT to map the Boolean assignments generated by the formal tool to a \red{set}\blue{sequence} of instructions.  
This mapping process uses the ISA of the target processor and is repeated for each clock cycle. 
The \textit{test case converter} also prepends this test case with another \red{set}\blue{sequences} of instructions that initialize and \blue{terminate} the processor\blue{'s simulation as the seed.}\red{ to a known state by resetting its memory and registers.} 
\red{Finally, the fuzzer uses the prepended test case as the seed.}

\input{Codes/pseudoalgo/overall}

\subsection{Putting It All Together}
As shown in Algorithm~\ref{alg:overall}, 
\ourtool{} first marks all the coverage points as uncovered. 
Then, we run a test case where we initialize all the registers and memory values of the DUT to zero using {\tt nop} instructions (Line 2). 
We then select an uncovered point ($p$) based on the selected strategy ($s$) (Line 8).
We then convert this $p$ into a corresponding {\it cover} property and invoke the formal tool to target this property (Line 9). 
The formal tool returns the Boolean assignment for each signal of the DUT for each clock cycle (Line 10), which is then converted to a sequence of instructions to be used as the seed of the fuzzer (Line 11). 
We then invoke the fuzzer on the DUT (Line 15). 
The fuzzer is continued to execute until the coverage rate of the fuzzer is less than that of the formal tool (Line 16); otherwise, we select a new uncovered point for the formal tool to target. This cycle continues until \ch{the target coverage is achieved or} the time limit is reached.

%% file: Codes/pseudoalgo/overall.tex
\begin{algorithm}[t]
\SetFuncSty{textsc}
\SetKwFunction{SwitchToFuzzer}{SwitchToFuzzer}
\SetKwFunction{SwitchToFormalTool}{SwitchToFormalTool}
\SetKwFunction{CallFuzzer}{CallFuzzer}
\SetKwFunction{InitMemAndRegToZero}{InitMemAndRegToZero}
\SetKwFunction{Fuzzer}{Fuzzer}
\SetKwFunction{MaxUncovd}{MaxUncovd}
\SetKwFunction{ModDep}{ModDep}
\SetKwFunction{BotTop}{BotTop}
\SetKwFunction{RandSel}{RandSel}
\SetKwFunction{SelectionStrategy}{SelectionStrategy}
\SetKwFunction{PropertyGenerator}{PropertyGenerator}
\SetKwFunction{TestCaseConverter}{TestCaseConverter}
\SetKwFunction{FormalTool}{FormalTool}
\SetKwFunction{FormalRate}{FormalRate}

\SetKwProg{Fn}{Function}{:}{}
\DontPrintSemicolon
\caption{\ourtool{}}\label{alg:overall}
\KwIn{
    $DUT$: design-under-test;\;
     \Indp \Indp $s$: point selection strategy;\; 
    $t_{limit}$: time limit of fuzzing;\;
    $tc$: target coverage;}
\KwOut{$n$: total coverage achieved;}
$t \gets 0$, $r_{fml} \gets 0$\;
Initialize registers and memory in DUT to zero\;
\While{$(n < tc)$ and $(t < t_{limit})$}{
    $r_{fml},seed,t \gets$ \SwitchToFormalTool($DUT$,$s$,$t$, $r_{fml}$)\;
    
    $n, t \gets $ \SwitchToFuzzer($DUT$,$t$,$r_{fml}$,$seed$,$t_{limit}$)\;
}
return $n$\;
\Fn{\SwitchToFormalTool{$DUT$, $s$, $t$, $r_{fml}$}}{
\tcc{select an uncovered point $p$ based on RandSel, MaxUncovd, BotTop, and ModDep strategies}
$p \gets $ \SelectionStrategy{$DUT$, $s$}

$cp_{rop} \gets $ \PropertyGenerator{$DUT$,$p$}\;
$Boolean\_assignment, r_{fml}  \gets$ \FormalTool{$DUT$,$cp_{rop}$}\;
$seed \gets$ \TestCaseConverter{$DUT$,$Boolean\_assignment$}\;

return $r_{fml}$, $seed$, $t$
}

\Fn{\SwitchToFuzzer{$DUT$,$t$,$r_{fml}$,$seed$, $t_{limit}$}}{
\Repeat{($r_{fuzz} < r_{fml}$) or ($t \geq t_{limit}$)}{
$n, t, r_{fuzz} \gets$ \Fuzzer{$DUT$,$seed$, $t$}\;
}
return $n, t$\;
}


\end{algorithm}

%% file: openarch/experiment.tex
\section{Evaluation}\label{sec:exp_results}

\input{Table/design_info}

\blue{We first evaluate how the rate of formal tool~($r_{fml}$) and fuzzer~($r_{fuzz}$) reflect the capability that the formal tool and the fuzzer explore the design spaces.}
We \red{first}\blue{then} evaluate \ourtool{} on five open-source processors from RISC-V~\cite{riscv_home} and OpenRISC~\cite{openrisc_home} instruction set architectures~(ISAs) to empirically select the best point selection strategy. We then compare \ourtool{} with the most recent hardware processor fuzzer~\cite{kandethehuzz} \red{in terms of}\blue{regarding} the coverage achieved and the vulnerabilities detected. \blue{Finally, we investigate the capability of \ourtool{} to cover points of different  coverage metrics.} We ran our experiments on a 32-core, 2.6 GHz Intel Xeon processor with 512 GB of RAM running Cent OS Linux release 7.9.2009.  

\subsection{Evaluation Setup} \label{cc:exp_setup}

\textbf{Benchmark selection.} Most commercial processors are protected intellectual property~(IP) and close-sourced.
\blue{Thus, we pick the three large (in terms of the number of gates) and  widely-used open-sourced processors: \rc{}~\cite{rocket_chip_generator}, \boom{}~\cite{boom}, and \cva{}~\cite{cva6} from the RISC-V ISA and \orth{}~\cite{or1200} and \morkx{}~\cite{mor1kx} from the OpenRISC ISA as the diverse set of benchmarks to evaluate \ourtool{}. Table~\ref{tab:design_info} lists the details of these processors.}
\red{Thus, the existing hardware fuzzers~\cite{rfuzz,hur2021difuzzrtl,canakci2021directfuzz,ragab_bugsbunny_2022,fuzzhwlikesw, kandethehuzz} used one or more of the following open-source processors in their evaluations: \rc{}~\cite{rocket_chip_generator}, \boom{}~\cite{boom}, \cva{}~\cite{cva6}, \texttt{BlackParrot}~\cite{blackparrot}, \texttt{Sodor cores}~\cite{sodor}, \orth{}~\cite{or1200}, and \morkx~\cite{mor1kx}.}
\red{Of these, we pick the top three largest (in terms of the number of gates and the number of lines of code) and  widely-used processors: \rc{}, \boom{}, and \cva{} from the RISC-V ISA and both \orth{} and \morkx{} from the OpenRISC ISA as the benchmarks to evaluate \ourtool{}.  
Table~\ref{tab:design_info} lists the details of these processors.}
\cva{} and \boom{} processors are complex than the \rc{}, \morkx{}, and \orth{}, with advanced micro-architectural features like out-of-order execution~(OoO) and single instruction-multiple data~(SIMD). 
\red{Additionally, the \cva{}, \boom{}, and \rc{} processors are larger than the \orth{} and \morkx{} processors, with around 10$\times$ more latches and gates, thereby providing a diverse set of benchmark designs to evaluate \ourtool{}.}

\noindent\textbf{Evaluation environment.} We use the popular and industry-standard Cadence \JG{}~\cite{jaspergold} and Synopsys \vcs{}~\cite{vcs} tools as the formal tool and the simulation tool, respectively. We use the branch coverage generated by \vcs{} to evaluate \ourtool{} because branch coverage is an important coverage metric in vulnerability detection~\cite{mockus2009test}.
\red{It also represents the number of control paths covered in the design-under-test~(DUT).}
Appendix~\ref{apd:cov_metric} details how we can evaluate \ourtool{} with other coverage metrics.
\ch{We use \textit{Chipyard}~\cite{chipyard} as a simulation environment for the RISC-V processors.}
For \ourtool{}'s fuzzer, we use the most recent hardware fuzzer for processors, \thehuzz{}~\cite{kandethehuzz}, as it achieves more coverage than prior hardware fuzzers, e.g., \difuzz{}~\cite{hur2021difuzzrtl}, and traditional random regression.  
To ensure a fair comparison, we constrain the environment of \JG{} to be the same as the hardware fuzzer's simulation environment~\cite{rfuzz, canakci2021directfuzz, hur2021difuzzrtl, kandethehuzz, ragab_bugsbunny_2022}. \blue{To prevent \JG{} from getting stuck while proving a property, we set a time limit on \JG{} to prove each property. 
We compute this time limit by using the time \JG{} spends on 30 random coverage points and applying survival analysis~\cite{kleinbaum1996survival} to calculate a time limit large enough to prove over $99\%$ of the points in the design.}
We set the rolling window size~($w$) as $100$ to calculate $r_{fuzz}$ using Equation~\ref{eq:r_fuzzing}. 
We ran each experiment for 72 hours and repeated it three times. 

\blue{\subsection{Evaluating Scheduling of Fuzzer and Formal Tool}\label{sec:rateAnalysis}}
\blue{We pick the coverage results of \ourtool{} on \cva{} to evaluate the coverage increment rate of the formal tool~($r_{fml}$) and the fuzzer~($r_{fuzz}$). The results are shown in Appendix~\ref{apd:eva_rfuzz_fml}.} 

\blue{\textbf{Evaluation on $r_{fml}$.} We evaluate how the $r_{fml}$ will quickly converge and reflect how fast the formal tool will explore the hard-to-reach spaces of the fuzzer in a design. The \textit{scheduler} updates $r_{fml}$ every time when \JG{} proves the reachability of an uncovered point. As shown in Figure~\ref{fig:rfml}, \ourtool{} switched to using \JG{} more than 300 times. 
The accumulated $r_{fml}$ only varies during the first 24 switches, after which the mean difference between the final $r_{fml}$ and the accumulated $r_{fml}$ is less than $5\%$. Hence, compared to the entire experiment, $r_{fml}$ only requires several samples to converge and can reflect the capability of space exploration of \JG{} in a design. The time limit on \JG{} also helps accelerate the convergence of $r_{fml}$.}

\blue{\textbf{Evaluation on $r_{fuzz}(w)$.} We evaluate how the rolling window size ($w$) affects the accuracy of estimating how well the fuzzer in \ourtool{} recently explored the design space.
The fuzzer in \ourtool{} executes $10$ test cases at a time; hence we analyze the $r_{fuzz}(w)$ for different values of $w$ in increments of 10. 
For each value of $w$, we plot the number of times the \textit{scheduler} will under-utilize the fuzzer and switch to a formal tool, as shown in Figure~\ref{fig:rfuzz}.
We consider the fuzzer under-utilized if the coverage of the fuzzer is temporarily stagnated but will cover new points through mutation of test cases if \textit{scheduler} had not switched to the formal tool.  
We can observe from Figure~\ref{fig:rfuzz} that when the window is set to $100$, $r_{fuzz}(w)$ can capture almost all coverage increments from the following test cases. Hence, we set the experiment $w$ to be $100$.}




\subsection{Evaluating Point Selection Strategies}\label{sec:covAnalysis}
To find the \red{optimal}\blue{most efficient} selection strategy for \ourtool{}, we now evaluate the three point selection strategies \bottop{}, \maxuncov{}, and \moddep{}, along with the \randsel{}, using two metrics: amount of coverage achieved and coverage speed.

\noindent\textbf{Coverage achieved.} 
Figure~\ref{fig:tot_branch} shows the branch coverage achieved by different selection strategies for all the processors. 
The difference between the coverage achieved by the four selection strategies is less than \minDiffCovStrats{}\% on all the processors except for the \cva{} processor, which has a difference of \cvaDiffCovStrats{}\%.
The difference in the coverage achieved by all the strategies decreases over time, and the coverage achieved will eventually converge.
\red{Still, they do differ in the rate of convergence. 
For example, in the case of the \boom{} processor, the coverage achieved by the four strategies is different, but after 72 hours, they converge.}
However, they still achieve an average of \avgPerMoreThe{}\% more coverage compared to \thehuzz{} and \avgPerMoreRR{}\% more coverage compared to random regression, across the five processors (see Section~\ref{sec:coverageAchieved}). 

\begin{figure}
    \centering
    \includegraphics[width=0.85\columnwidth]{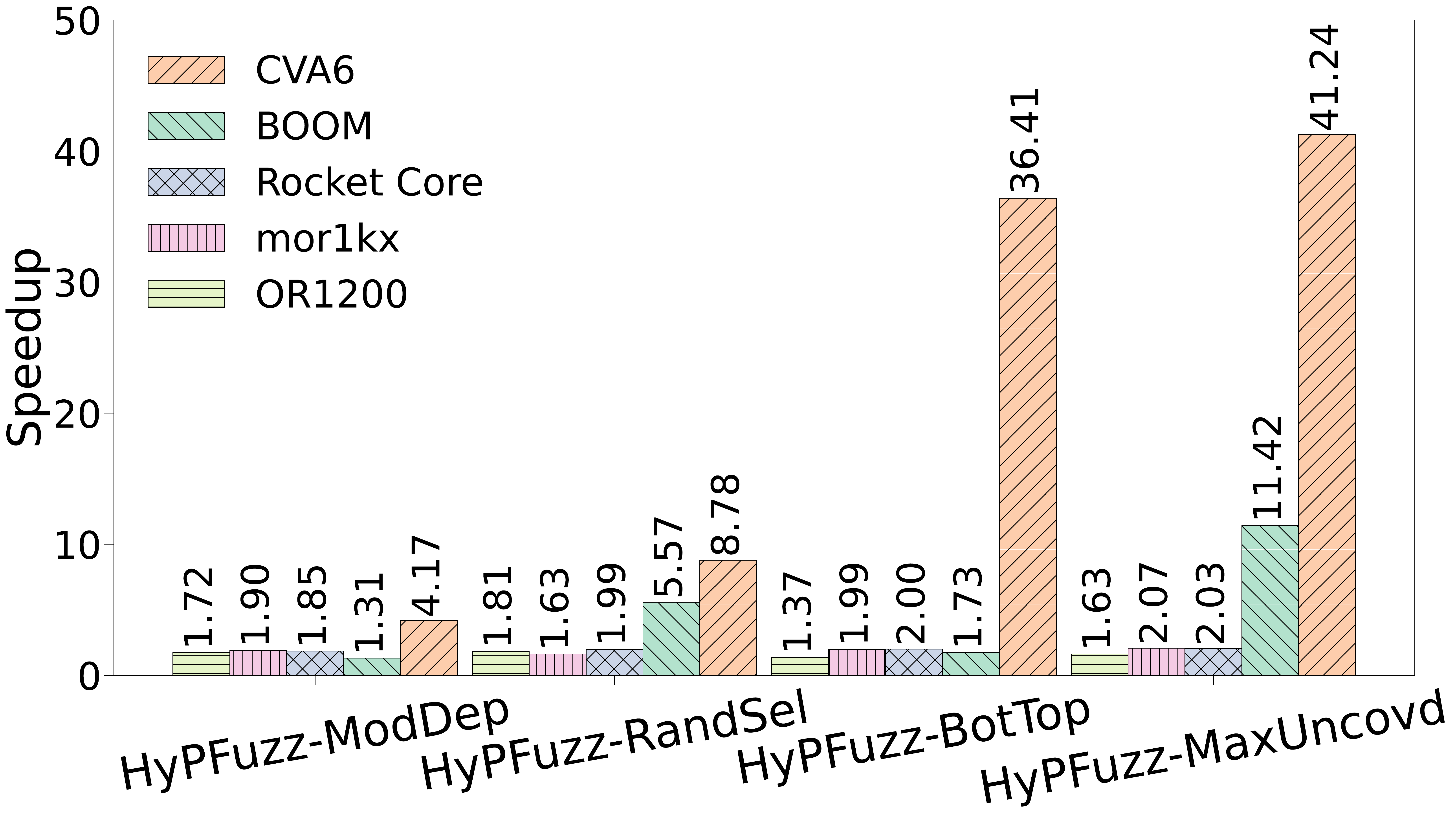}
    \caption{Speedup of \ourtool{} over 
    \thehuzz{}~\cite{kandethehuzz}.}
    \label{fig:branch_rate}
\end{figure}

\noindent\textbf{Coverage speed} of the four selection strategies of \ourtool{} on the five processors is shown in Figure~\ref{fig:branch_rate}. 
The \maxuncov{} selection strategy is the fastest for all processors except for the \orth{}, where the \randsel{} strategy performs the best.
The reason why \randsel{} strategy is the fastest for the \orth{} processor is because of its smaller size (around 25K gates), and \randsel{} has a faster selection time compared to other strategies. 
However, this is not the case for the other processors, especially the larger processors, such as \cva{} and \boom{}, which have several hundred thousand gates and more microarchitectural features. For these designs, the order of point selection has a more dominant effect on coverage speed than the time taken to select the point. 
Based on these analyses, we select the \maxuncov{} strategy as our \red{optimal }point selection strategy for \ourtool{}. All further evaluations will use the \maxuncov{} strategy unless stated otherwise.

\subsection{Coverage Achieved}\label{sec:coverageAchieved}
We now evaluate the capability of random regression, \thehuzz{},  and \ourtool{} in achieving coverage.
Across the five processors, \ourtool{} achieves \textbf{\avgPerMoreRR{}\%} more coverage than random regression and is \textbf{\aveCovSpeedupvsRandreg{}}$\bm{\times}$ faster than random regression after fuzzing for 72 hours, as seen in Figure~\ref{fig:tot_branch}. 
Also, \ourtool{} achieves \textbf{\avgPerMoreThe{}\%} more coverage than \thehuzz{} and is  \textbf{\aveCovSpeedup}$\bm{\times}$ faster than \thehuzz{} after running for the same 72 hours, as seen in Figure~\ref{fig:tot_branch} and Figure~\ref{fig:branch_rate}\blue{, respectively}.

\begin{figure}
    \centering
    \includegraphics[width=0.85\columnwidth]{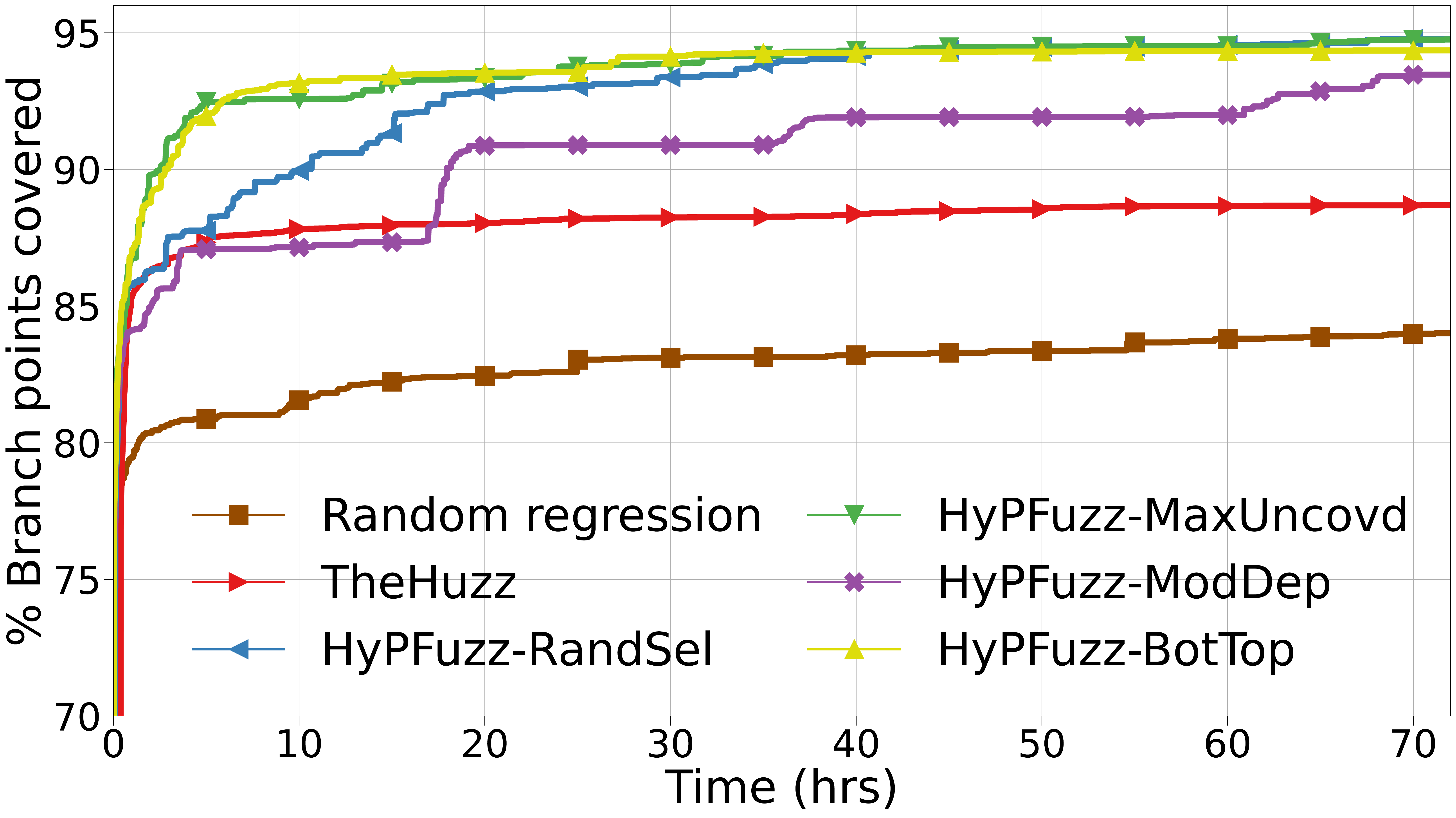}
    \caption{Total branch points covered by random regression, \thehuzz{}~\cite{kandethehuzz}, and \ourtool{} on \cva{}~\cite{cva6}.}
    \label{fig:cva6_tot_exp}
\end{figure}

The \cva{} and \boom{} processors are complex with advanced OoO and SIMD features. \blue{Hence, \ourtool{} achieves substantially higher coverage than \thehuzz{} on them. Figure~\ref{fig:cva6_tot_exp} shows that \ourtool{} achieves the highest branch coverage improvement of \maxPerMoreTh{}\% on \cva{} compared to \thehuzz{}.} 
Thus, the coverage speed of TheHuzz is less for these processors.
However, \ourtool{} leverages the \red{optimal}\blue{efficient} \textit{point selector} to generate \red{optimal }seeds and maximize the coverage achieved by its fuzzer. This results in a speedup of \ratecva{}$\times$ on \cva{} and \rateboom{}$\times$ on \boom{} compared to \thehuzz{}. 
On the other hand, the \rc{}, \morkx{}, and \orth{} do not have advanced OoO and SIMD features. 
However, \ourtool{} is still at least \minSpeedupThreeprocs{}$\times$ faster than \thehuzz{}. This is because the dynamic \textit{scheduler} adapts to the complexity of the DUT using the values of $r_{fml}$ and $r_{fuzz}$, scheduling the fuzzer and the formal tool \red{optimally}\blue{efficiently}. 

\red{\input{Table/exp_stat_anal}}

\red{\noindent\textbf{Statistical significance.}
To analyze the statistical significance of our evaluations, we compute the p-value of the Mann-Whitney \textit{U} test~\cite{utest} and Vargha Delaney's A12 measure~\cite{a12}, which are widely used to evaluate fuzzers~\cite{kandethehuzz,hur2021difuzzrtl, fuzzhwlikesw}.}

\red{A p-value of less than 0.05 indicates that two distributions are statistically different~\cite{hur2021difuzzrtl}.
From Table~\ref{tab:stat_anal} it can be seen that the coverage results are statistically different for all the processors other than the \orth{} processor. 
The \orth{} processor has $p > 0.05$ because the coverage results of \ourtool{} and \thehuzz{} are very close, with 
\ourtool{} achieving only \covPerMoreOrthHyPTheH{}\% more coverage than \thehuzz{}. 
Also, the A12 measure indicates that \ourtool{} achieves more  coverage over \thehuzz{} for all the processors.}

In summary, \ourtool{} achieved more coverage and is faster than random regression and \thehuzz{} on all five processors. 
\ourtool{} is faster when fuzzing processors, including the ones with complex micro-architectural features. 
However, as the complexity and size of the processor grow, the \textit{point selector} and the \textit{scheduler} become more effective, resulting in more speedup achieved by \ourtool{}.

\input{Table/bug_table}

\subsection{Vulnerabilities Detected}\label{sec:bugsDetected}
While the coverage achieved sheds light on the extent of the DUT verified, it is not a direct measure of a tool's ability to detect vulnerabilities. Hence, we evaluate \ourtool{}'s ability to detect vulnerabilities in real-world processors.  

\noindent\textbf{Vulnerability detection strategy} of \ourtool{} is the same as that of many existing fuzzers~\cite{hur2021difuzzrtl, ragab_bugsbunny_2022, canakci2021directfuzz, kandethehuzz}. For the same test case, we  output the architecture states of both the DUT and the golden reference model~(GRM). 
The architecture state includes the value of general purpose registers (GPRs), control \red{state}\blue{and status} registers (CSRs), instructions committed, and exceptions triggered. Any mismatch between the architecture states indicates a potential vulnerability in the DUT or the GRM. In our experiment, we use the RISC-V ISA emulator, \spike{}~\cite{spike}, as the GRM for \rc{}, \boom{}, and \cva{} processors and the OpenRISC ISA emulator, \textit{or1ksim}~\cite{or1ksim}, for \orth{} and \morkx{} processors.  

\noindent Apart from detecting all the vulnerabilities detected by the most recent fuzzer~\cite{kandethehuzz}, \ourtool{} detected \textbf{three new vulnerabilities},  as listed in Table~\ref{tb:b_list_v1}. 
\red{We briefly describe the three new vulnerabilities (see Appendix~\ref{apd:NewBugs} for more details).}

\noindent\begin{bug_ns}[\label{v1}] \blue{is in the memory control unit of \cva{} processor and} is similar to an out-of-bounds memory access vulnerability in software programs~\cite{serebryany2012addresssanitizer}.\red{, which can lead to failure of memory isolation protections and leak data~(common weakness enumeration~} 
\blue{According to the \riscv{} specification~\cite{riscv_home}, a processor must raise an exception when operations try to access data at invalid memory addresses. However, \cva{} will not raise exceptions in the same situation. We detected this vulnerability as a mismatch when \spike{} raised an exception for such operations and interrupted the program, whereas \cva{} continued to execute the program. Because operating systems usually use such exceptions to protect isolated executable memory space, missing them can allow an attacker to access data from all of memory~}(CWE-1252~\cite{cwe1252}).
\end{bug_ns}

\noindent\begin{bug_ns}[\label{v2}] \blue{is located in the decode stage of \cva{} processor and} is similar to an undefined behavior vulnerability in a software program~\cite{unbehave}.
\red{The decoder in the \cva{} processor permits execution of \texttt{MULH} instructions with invalid destination registers.}
\blue{According to the \riscv{} specification~\cite{riscv_home}, the decoder should throw an illegal instruction exception when the destination register~(\texttt{rd}) of instruction \texttt{MULH} is the same as the first~(\texttt{rs1}) or second~(\texttt{rs2}) source register. This specification will reduce the utilization of multiplier units and increase the performance of processors. The vulnerability is that the decoder in \cva{} allows the \texttt{rd} of \texttt{MULH} to share the same register as \texttt{rs1} or \texttt{rs2}. This means the design of \cva{} violates the ISA specification. We detected this vulnerability when \ourtool{} generated a test case containing \texttt{MULH} with the same \texttt{rd}, \texttt{rs1}, and \texttt{rs2}. \spike{} threw an illegal instruction exception, whereas \cva{} executed the instruction. This vulnerability can result in a potential performance bottleneck when executing applications with heavy multiplier operations, such as machine learning~(CWE-440~\cite{cwe440}).} 
\end{bug_ns}

\noindent\begin{bug_ns}[\label{v3}] is a cross-modular vulnerability where the read logic in the CSR module enables access to undefined hardware performance counters~(HPCs), resulting in unknown values when these HPCs are accessed. 
\blue{It is similar to an undefined behavior in a software program~\cite{unbehave}. 
\cva{} has implemented 14 HPCs for recording various hardware behaviors, and its CSR module is responsible for reading the value of an HPC based on requests from the operating system. However, the reading logic in the CSR module enables access to 32 HPCs, which causes X-propagation when reading nonexistent HPCs. We detected this vulnerability when \spike{} returns regular numbers while \cva{} returns X~(unknown) values. This vulnerability will cause potential issues during synthesis and fail the functions of HPC (CWE-1281~\cite{cwe1281}).}
\end{bug_ns}

Vulnerabilities~\ref{v2} and~\ref{v3} resulted in two new common vulnerabilities and exposures~(CVE) entries, \texttt{CVE-2022-33021} and \texttt{CVE-2022-33023}.

\noindent\textbf{Comparison with \thehuzz{}.}
In addition to the three new vulnerabilities, \ourtool{} detected all 11 vulnerabilities reported by \thehuzz{} as listed in Table~\ref{tb:b_list_v1}. 
The speedup column shows how fast \ourtool{} is compared to \thehuzz{} in terms of both run-time and number of instructions. 
It can be seen that the speed of vulnerability detection follows the same trend as the coverage speeds.
\ourtool{} detects vulnerabilities in \cva{} at faster than the vulnerabilities in \rc{}, \morkx{}, and \orth{} processors.  
However, \thehuzz{} detected some vulnerabilities faster than \ourtool{}. 
This is because the number of instructions required to trigger them is too low (< 1000 instructions) (i.e., measure the coverage and mutate test cases). 
For example, V9 and V13 require less than one hundred instructions to trigger. 
On an average, \ourtool{} detects vulnerabilities  \textbf{\avebugtimerate{}}$\bm{\times}$ faster than \thehuzz{} in terms of run time and uses \textbf{\avebuginstrate}{}$\bm{\times}$   fewer number of instructions to trigger them.

\subsection{Evaluation on Condition and FSM Coverage Metrics}\label{sec:complexCovMetrics}
\blue{In this section, we demonstrate the compatibility of \ourtool{} in achieving coverage with different coverage metrics such as \textit{condition} and \textit{finite-state machine}~(FSM) coverage.
\ourtool{} uses its \textit{property generator} to generate the \textit{cover} properties for the points of them as discussed in Appendix~\ref{apd:cov_metric}.
Figure~\ref{fig:cond_fsm} shows the coverage achieved by \ourtool{} and \thehuzz{} on \cva{} processor when run for $72$ hours and repeated three times.
\ourtool{} achieves an average of $\CondAveCovImprv{}\%$ more \textit{condition} coverage and $\FSMAveCovImprv{}\%$ more \textit{FSM} coverage compared to \thehuzz{} hence demonstrating that \ourtool{} is capable of improving coverage for different types of coverage metrics.
}




\blue{}

\begin{figure}
    \centering
    \includegraphics[width=0.85\linewidth]{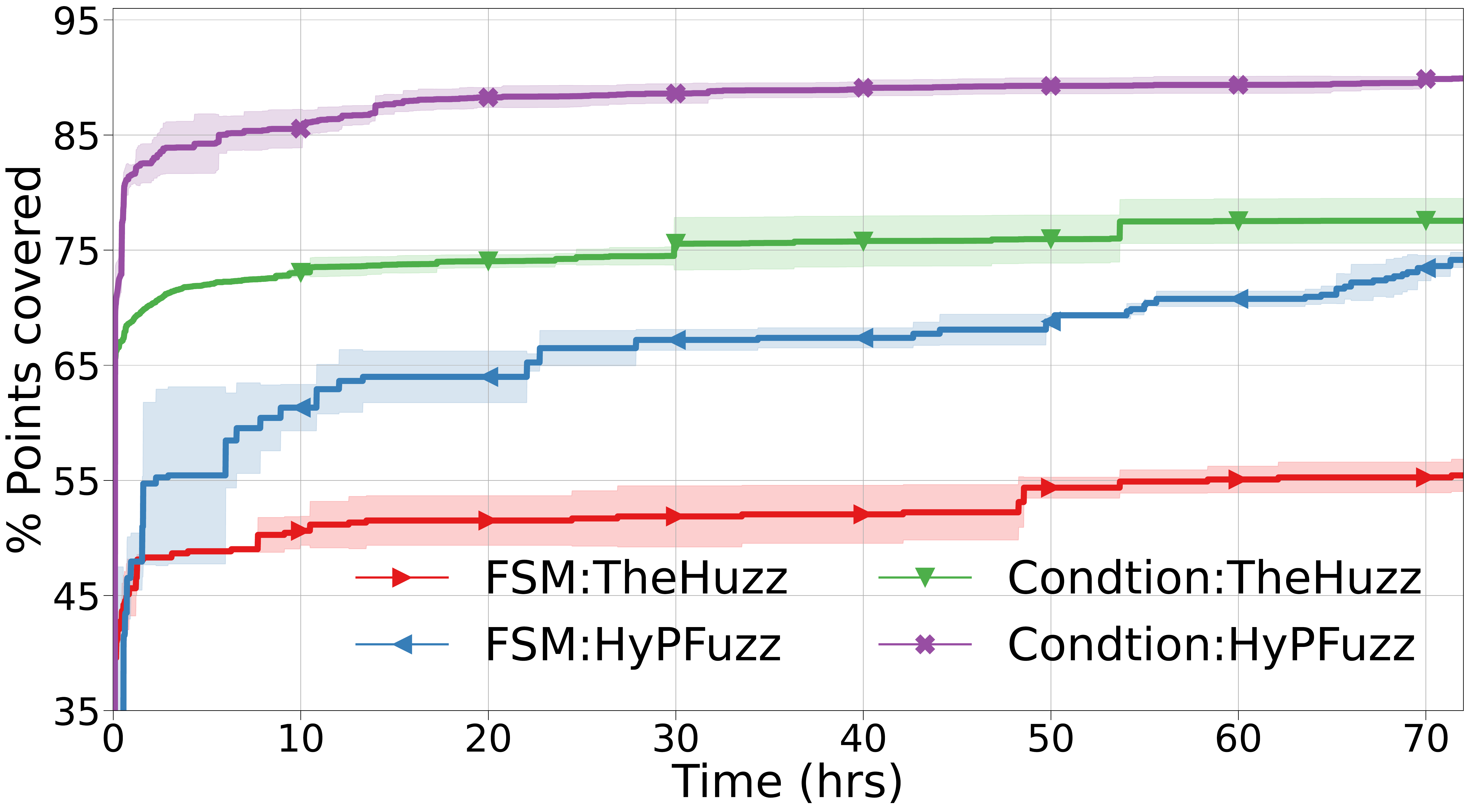}
    \caption{Total condition points and FSM points covered by \thehuzz{}~\cite{kandethehuzz} and \ourtool{}.}
    \label{fig:cond_fsm}
\end{figure}



%% file: Table/design_info.tex
\begin{table}[]
\centering
\caption{Benchmarks used in our study.
}
\label{tab:design_info}
\resizebox{\columnwidth}{!}{%
\begin{tabular}{|c|c|c|c|c|c|c|}
\hline
\textbf{Processor} & \textbf{\# \red{l}\blue{L}atches} & \textbf{\# \red{g}\blue{G}ates} & \textbf{\# \red{b}\blue{B}ranch points} & \textbf{OoO} & \textbf{SIMD} & \blue{\textbf{Time limit~(sec)}}\\ \hline
\orth{}~\cite{or1200}  & $3.08\times 10^3$ & $2.48 \times10^4$ & $7.00\times10^2$ & \tikzxmark{}   & \tikzxmark{} & \blue{$4.80\times10^2$}\\ \hline
\morkx{}~\cite{mor1kx} & $5.29\times 10^3$ & $4.64\times10^4$ & $1.34\times10^3$ & \tikzxmark{}   & \tikzxmark{} & \blue{$6.98\times10^2$}\\ \hline
\cva{}~\cite{cva6}  & $2.48\times10^4$ & $4.63\times10^5$ & $9.53\times10^3$ & \tikzcmark{} & \tikzcmark{} & \blue{$3.80\times10^3$}\\ \hline
\rc{}~\cite{rocket_chip_generator} & $1.64\times10^5$ & $9.24\times10^5$ & $1.33\times10^4$ & \tikzxmark{}   & \tikzxmark{} & \blue{$2.71\times10^3$}\\ \hline
\boom{}~\cite{boom}  & $1.99 \times 10^5$ & $ 1.26\times10^6 $ & $2.42\times10^4$ & \tikzcmark{} & \tikzxmark{} & \blue{$3.83\times10^3$}\\ \hline
\end{tabular}%
}
\end{table}

%% file: Table/bug_table.tex

\begin{table*}[!bth]
\centering
\caption{Vulnerabilities detected by \ourtool{}. N.A. denotes ``Not Applicable.'' }
\resizebox{\textwidth}{!}{%
\begin{tabular}{|c|l|c|c|ccl|ccl|}
\hline
\multirow{2}{*}{\textbf{Processor}} & \multicolumn{1}{c|}{\multirow{2}{*}{\textbf{Vulnerability Description}}} & \multirow{2}{*}{\textbf{CWE}} & \multirow{2}{*}{\begin{tabular}[c]{@{}c@{}}\textbf{New?}\end{tabular}} & \multicolumn{3}{c|}{\textbf{\# Instruction\blue{s}}}                          & \multicolumn{3}{c|}{\textbf{Time~(sec)}}                               \\ \cline{5-10} 
                          & \multicolumn{1}{c|}{}                                 &                      &                                                              & \multicolumn{1}{c|}{\textbf{\thehuzz{}}} & \multicolumn{1}{l|}{\textbf{\ourtool{}}} & \textbf{Speedup} & \multicolumn{1}{c|}{\textbf{\thehuzz{}}} & \multicolumn{1}{l|}{\textbf{\ourtool{}}} & \textbf{Speedup} \\ \hline
\multirow{7}{*}{\cva{}~\cite{cva6}}      & \begin{tabular}[c]{@{}l@{}}V1: Missing exceptions when \\ accessing invalid addresses. \end{tabular} & CWE-1252 & \tikzcmark{}  & \multicolumn{1}{c|}{$N.A.$} & \multicolumn{1}{c|}{$2.67 \times 10^1$} & \multicolumn{1}{c|}{$N.A.$} & \multicolumn{1}{c|}{$N.A.$}  & \multicolumn{1}{c|}{$6.67 \times 10^1$} & \multicolumn{1}{c|}{$N.A.$}                         \\ \cline{2-10} 
                          & \begin{tabular}[c]{@{}l@{}}V2: Incorrect decoding logic \\ for multiplication instructions. \end{tabular} & CWE-440 & \tikzcmark{} & \multicolumn{1}{c|}{$N.A.$}     & \multicolumn{1}{c|}{$9.80 \times 10^4$} &    \multicolumn{1}{c|}{$N.A.$}      & \multicolumn{1}{c|}{$N.A.$}     & \multicolumn{1}{c|}{$4.20 \times 10^3$} & \multicolumn{1}{c|}{$N.A.$}                          \\ \cline{2-10} 
                          & \begin{tabular}[c]{@{}l@{}}V3: Returning X-value when \\ access unallocated CSRs. \end{tabular} & CWE-1281 & \tikzcmark{} & \multicolumn{1}{c|}{$N.A.$} & \multicolumn{1}{c|}{$2.08 \times 10^3$} & \multicolumn{1}{c|}{$N.A.$}      & \multicolumn{1}{c|}{$N.A.$}     & \multicolumn{1}{c|}{$1.70 \times 10^2$} &  \multicolumn{1}{c|}{$N.A..$}         \\ \cline{2-10} 
                          & \begin{tabular}[c]{@{}l@{}}V4: Failure to detect cache \\ coherency violation. \end{tabular}                 & CWE-1202                 & \tikzxmark{}  & \multicolumn{1}{c|}{$1.72 \times 10^5$}  & \multicolumn{1}{c|}{$2.15 \times 10^5$}  & \multicolumn{1}{c|}{$0.80\times $} & \multicolumn{1}{c|}{$6.50 \times 10^3$} & \multicolumn{1}{c|}{$8.40 \times 10^3$} &    \multicolumn{1}{c|}{$0.77\times $}                      \\ \cline{2-10} 
                          & \begin{tabular}[c]{@{}l@{}}V5: Incorrect decoding logic \\ for the FENCE.I instruction. \end{tabular} & CWE-440 & \tikzxmark{} & \multicolumn{1}{c|}{$1.36 \times 10^4$} & \multicolumn{1}{c|}{$1.08 \times 10^4$} & \multicolumn{1}{c|}{$1.26\times $} & \multicolumn{1}{c|}{$1.68 \times 10^3$}     & \multicolumn{1}{c|}{$3.08 \times 10^2$} & \multicolumn{1}{c|}{$5.45\times $}      \\ \cline{2-10} 
                          & \begin{tabular}[c]{@{}l@{}}V6: Incorrect exception type \\ in instruction queue. \end{tabular} & CWE-1202 & \tikzxmark{} & \multicolumn{1}{c|}{$4.02 \times 10^4$} & \multicolumn{1}{c|}{$7.40 \times 10^3$} & \multicolumn{1}{c|}{$5.43\times $} & \multicolumn{1}{c|}{$2.54 \times 10^3$}     & \multicolumn{1}{c|}{$2.92 \times 10^2$} & \multicolumn{1}{c|}{$8.69\times $}                          \\ \cline{2-10} 
                          & \begin{tabular}[c]{@{}l@{}}V7: Missing exceptions for \\ some \textit{illegal} instructions. \end{tabular}              & CWE-1242                 & \tikzxmark{}                                                                   & \multicolumn{1}{c|}{$1.81 \times 10^6$}     & \multicolumn{1}{c|}{$1.43 \times 10^5$} & \multicolumn{1}{c|}{$12.66\times $} & \multicolumn{1}{c|}{$5.67 \times 10^4$}     & \multicolumn{1}{c|}{$6.41 \times 10^3$} & \multicolumn{1}{c|}{$8.85\times $} \\ \hline
\rc{}~\cite{rocket_chip_generator} & \begin{tabular}[c]{@{}l@{}}V8: Instruction commit count \\ not increased when \texttt{EBREAK}. \end{tabular} & CWE-1201 & \tikzxmark{}                                                                   & \multicolumn{1}{c|}{$7.76 \times 10^2$}     & \multicolumn{1}{c|}{$1.07 \times 10^2$} & \multicolumn{1}{c|}{$7.25\times $}   & \multicolumn{1}{c|}{$1.19 \times 10^2$}     & \multicolumn{1}{c|}{$6.67 \times 10^1$} & \multicolumn{1}{c|}{$1.78\times $}                          \\ \hline
\multirow{3}{*}{\morkx{}~\cite{mor1kx}}    & \begin{tabular}[c]{@{}l@{}}V9: Incorrect implementation of \\ the \texttt{carry} flag generation. \end{tabular} & CWE-1201 & \tikzxmark{}                              & \multicolumn{1}{c|}{$20$}      &  \multicolumn{1}{c|}{$20$} & \multicolumn{1}{c|}{$1.00\times$}     & \multicolumn{1}{c|}{$10$}    & \multicolumn{1}{c|}{$10$} &    \multicolumn{1}{c|}{$1.00\times$}                   \\ \cline{2-10} 
                          & \begin{tabular}[c]{@{}l@{}}V10: Missing access checking \\ for privileged register. \end{tabular} & CWE-1262 & \tikzxmark{}                 & \multicolumn{1}{c|}{$4.46 \times 10^5$} & \multicolumn{1}{c|}{$2.30 \times 10^5$} & \multicolumn{1}{c|}{$1.94\times $} & \multicolumn{1}{c|}{$1.97 \times 10^4$}      &    \multicolumn{1}{c|}{$8.20 \times 10^3$} & \multicolumn{1}{c|}{$2.40\times $}                        \\ \cline{2-10} 
                          & \begin{tabular}[c]{@{}l@{}}V11: Incomplete implementation of \\ the \texttt{EEAR} register write logic. \end{tabular} & CWE-1199 & \tikzxmark{}                   & \multicolumn{1}{c|}{$1.12 \times 10^5$}      & \multicolumn{1}{c|}{$2.13 \times 10^5$} & \multicolumn{1}{c|}{$0.53\times $} & \multicolumn{1}{c|}{$4.89 \times 10^3$}      & \multicolumn{1}{c|}{$7.60 \times 10^3$} & \multicolumn{1}{c|}{$0.64\times $}                           \\ \hline
\multirow{3}{*}{\orth{}~\cite{or1200}}    & \begin{tabular}[c]{@{}l@{}}V12: Incorrect forwarding logic \\ for the \texttt{GPR0}. \end{tabular}  & CWE-1281 & \tikzxmark{}  & \multicolumn{1}{c|}{$1.05\times10^3$}      & \multicolumn{1}{c|}{$1.23 \times 10^3$} & \multicolumn{1}{c|}{$0.85\times$}  & \multicolumn{1}{c|}{$1.12\times10^2$}      &    \multicolumn{1}{c|}{$60$} &    \multicolumn{1}{c|}{$1.87\times$}      \\ \cline{2-10} 
                          & \begin{tabular}[c]{@{}l@{}}V13: Incorrect \texttt{overflow} logic \\ for \texttt{MSB} \& \texttt{MAC} instructions. \end{tabular} & CWE-1201 & \tikzxmark{}   & \multicolumn{1}{c|}{$47$}      & \multicolumn{1}{c|}{$73$} &  \multicolumn{1}{c|}{$0.64\times$}  & \multicolumn{1}{c|}{$1.25\times10^1$}      & \multicolumn{1}{c|}{$5.05\times10^1$} & \multicolumn{1}{c|}{$0.25\times$}                           \\ \cline{2-10} 
                          & \begin{tabular}[c]{@{}l@{}}V14: Incorrect generation of \\ \texttt{overflow} flag. \end{tabular} & CWE-1201 & \tikzxmark{} & \multicolumn{1}{c|}{$2.21\times10^4$}      & \multicolumn{1}{c|}{$1.78 \times 10^4$} & \multicolumn{1}{c|}{$1.24\times$}   & \multicolumn{1}{c|}{$1.34\times10^3$}      & \multicolumn{1}{c|}{$6.72\times10^2$} & \multicolumn{1}{c|}{$1.99\times$}  \\ \hline
\end{tabular}
}
\label{tb:b_list_v1}
\end{table*}

%% file: openarch/relatedwork.tex
\section{Related Works}\label{sec:related_works}
\red{We now describe the existing hardware fuzzers, highlight their limitations, and highlight the advantages of \ourtool{}.} 

\blue{\textbf{Hardware fuzzers.} \textbf{\rfuzz}{~\cite{rfuzz}} uses mux-toggle coverage as feedback to fuzz hardware and was one of the first attempts towards creating hardware fuzzers. The mux-toggle coverage covers the activity in the select signals of muxes in the DUT.}
\blue{\textbf{\textit{DirectFuzz}}~\cite{canakci2021directfuzz} leverages the mux-toggle coverage and allocates more mutation energy to test cases that achieve coverage closed to a manually selected model, hence achieving faster coverage. However, the mux-toggle metric does not scale well to large designs like \boom{} and \cva{} due to instrumentation overhead.}
\blue{\textbf{\difuzz{}}~\cite{hur2021difuzzrtl} addresses the overhead limitation of \rfuzz{} by developing a new coverage metric, control-register coverage. It uses all possible combinations of values of all the registers driving the selection logic of MUX. However, this coverage metric does not assign coverage points to many registers as well as combinational logic, thereby missing many security-critical vulnerabilities~\cite{kandethehuzz}.}
\blue{Therefore, \textbf{\thehuzz{}}~\cite{kandethehuzz}  uses code coverage metrics (branch, condition coverage, etc.) as feedback information to guide the fuzzer. \thehuzz{} cannot verify the DUT efficiently as it is increasingly difficult to cover the coverage points in hard-to-reach regions of hardware~(see Section~\ref{sec:fuzzer_limitations}). The coverage achieved by this processor is only 63\%.}
\blue{\textbf{\textit{Fuzzing hardware like software}}~\cite{fuzzhwlikesw}, unlike other hardware fuzzers, first translates a hardware design into a software model, then uses the coverage-guided software fuzzers to fuzz hardware. However, the software model does not support all the hardware constructs like latches and floating wires~\cite{kandethehuzz}.}
\red{\noindent\textbf{\rfuzz{}}~\cite{rfuzz} uses mux-toggle coverage as feedback to fuzz hardware and was one of the first attempts towards creating hardware fuzzers. The mux-toggle coverage covers the activity in the select signals of muxes in the DUT. However, it does not scale well to large designs like the \boom{} processor~\cite{boom} with $1.75 \times 10^{4}$ lines of \chisel{}~\cite{bachrach2012chisel} code, as reported in \difuzz{}~\cite{hur2021difuzzrtl}.}

\red{\noindent\textbf{\difuzz{}}~\cite{hur2021difuzzrtl} addresses the overhead limitation of \rfuzz{} by using a new self-developed coverage metric, control-register coverage, as feedback. Control-register coverage uses all possible combinations of values of all the registers driving the selection logic of MUX. 
However, this coverage metric does not assign coverage points to many  registers as well as combinational logic, thereby missing many security-critical vulnerabilities~\cite{kandethehuzz}.}

\red{\noindent\textbf{\textit{DirectFuzz}}~\cite{canakci2021directfuzz} generates test inputs to maximize the coverage of a manually selected module instead of the whole design. It uses the coverage metrics of \rfuzz{} and prioritizes input tests that increase coverage close to the selected module. 
Even though \textit{DirectFuzz} can achieve faster coverage than \rfuzz{} for the selected module, they did not fuzz the entire processor, thereby missing software-exploitable vulnerabilities.} 

\red{\noindent\textbf{\textit{BugsBunny}}~\cite{ragab_bugsbunny_2022}, akin to \textit{DirectFuzz}, fuzzes specific target signals in the DUT. 
This fuzzer is only useful in verifying target signals of single-module register-transfer level (RTL) designs and  does not support designs with multiple modules like processors.} 

\red{\noindent\textbf{\thehuzz{}}~\cite{kandethehuzz}  uses code coverage metrics (statement, branch, condition, FSM, and toggle coverage) as feedback information to guide the fuzzer. \thehuzz{} cannot verify the DUT entirely as it is increasingly difficult to cover the coverage points in hard-to-reach regions of hardware~(see Section~\ref{sec:fuzzer_limitations}). The coverage rate achieved by this processor is only 63\%.}

\red{\noindent\textbf{\textit{Fuzzing hardware like software}}~\cite{fuzzhwlikesw} is a novel hardware fuzzer that, unlike other hardware fuzzers, first translates a hardware design into a software model, then uses the coverage-guided software fuzzers to fuzz hardware. 
It uses \verilator{}~\cite{verilator}, an open-source hardware simulator, to translate the hardware into a software model which does not support all the hardware constructs like latches and floating wires~\cite{kandethehuzz}.}  

In summary, existing hardware fuzzers suffer from one or more of the following limitations: \blue{they} (i)~\red{they }cannot scale to large designs, (ii)~cannot capture all hardware behavior \blue{in RTL}, thereby missing vulnerabilities, (iii)~are not designed to fuzz the entire DUT, or (iv)~do not support all the constructs like latches and floating wires. 

\blue{\noindent\textbf{Hybrid techniques.} Existing software hybrid techniques combine fuzzers with symbolic execution~\cite{zhu2022fuzzing,godefroid2005dart,godefroid2008automated,pak2012hybrid,pham2016model,zhao2019send,stephens2016driller}. Fuzzers leverage symbolic execution's constraint-solving capabilities to explore \textit{deep} execution paths. Meanwhile, symbolic execution utilizes concrete test cases generated by fuzzers to mitigate scalability issues~\cite{kadron2023fuzzing}. Therefore, most hybrid fuzzers run symbolic execution in parallel to keep calculating paths covered by test cases and also use symbolic execution to generate new test cases when the original program terminates normally~(halt) or abnormally~(e.g., crash)~\cite{godefroid2005dart,godefroid2008automated,pak2012hybrid,pham2016model,zhao2019send}. \textbf{\textit{Driller}}~\cite{stephens2016driller} invokes symbolic execution when the fuzzer cannot achieve coverage after a pre-defined time threshold. Moreover, \textit{Driller} will select an unexplored path close to the paths covered by test cases. \textbf{\textit{DART}}~\cite{godefroid2005dart} applies a depth-first search along the path tree, and \textbf{\textit{SAGE}}~\cite{godefroid2008automated} selects paths under the paths explored by previous inputs with maximal coverage increment. \textbf{\textit{Pak's}} strategy~\cite{pak2012hybrid} profiles unexplored paths and assigns difficult ones with higher priority to be selected, and \textbf{\textit{Digfuzz}}~\cite{zhao2019send} applies Monte Carlo methods~\cite{robert2009monte} to quantify the difficulty of unexplored paths with probability.}

\blue{Compared to software hybrid fuzzers, \ourtool{}: (i)~generates SVA properties and invokes formal tools, ensuring compatibility with various coverage metrics for different security and verification purposes rather than just path, (ii)~does not require formal tools to check the test cases generated by the fuzzers, utilizing formal tools only when necessary, (iii)~develops multiple uncovered point selection strategies based on computer architecture and hardware design routines~\cite{hennessy2011computer,shen2013modern}.}

Apart from these fuzzing techniques, researchers have combined various formal verification techniques with either random regression~\cite{ho2000smart,kolbi2001symbolic} or fuzzing~\cite{li2021symbolic} in the context of test generation, targeting faults in hardware rather than security vulnerabilities, which is the focus of \ourtool{}. Also, none of these techniques scale to large, real-world designs.

In contrast, \ourtool{}: (i)~is scalable to large real-world processors, (ii)~captures all activity in the hardware details, such as FSMs and combinational logic, (iii)~fuzzes the entire processor, (iv)~uses industry-standard hardware simulators that support all hardware constructs and has the ability to detect vulnerabilities.

\red{Also, \ourtool{} achieved \aveCovSpeedup{}$\times$ faster coverage than the most recent hardware fuzzer~\cite{kandethehuzz} and detected vulnerabilities \avebugtimerate{}$\times$ faster. 
\ourtool{} detected three new vulnerabilities that led to two common vulnerabilities and exposures~(CVEs).}

%% file: openarch/discussion.tex
\section{Discussion}\label{sec:discuss}

\red{\noindent \textbf{Scalability of \ourtool{}.} 
The fuzzer and the formal tool in \ourtool{} use industry-standard electronic design automation~(EDA) tools, Synopsys \vcs{}~\cite{vcs} and Cadence \JG{}~\cite{jaspergold}, respectively, capable of handling commercial-scale designs. 
Previously researchers have shown that formal tools cannot scale well when targeting security vulnerabilities in processors~\cite{dessouky2019hardfails}. 
We tackle this problem by limiting the formal tool's usage only to target the \textit{cover} properties. 
Our results also indicate that \ourtool{} can successfully fuzz large processors, such as the million-gate \boom{} and 300K-gate \cva{}~\cite{cva6} processors.}

\noindent\blue{\textbf{Utilization of point selection strategies.} \ourtool{} specifically adopts the \maxuncov{} strategy to select uncovered points since it shows the highest coverage speed compared to the other strategies. However, we can modify \ourtool{} to switch from them dynamically. We can assign weight to each strategy as the probability of it being selected by \ourtool{}. We then use evolutionary algorithms, such as particle swarm optimization algorithm~\cite{shi1998modified,lyu2019mopt}, to dynamically update the weights for fast and high coverage.}

\red{\noindent\textbf{Availability of golden reference model (GRM).} \ourtool{} relies on a GRM to detect vulnerabilities, similar to other existing hardware fuzzers such as~\cite{hur2021difuzzrtl,kandethehuzz,ragab_bugsbunny_2022}. 
ISA simulators or software models of the processors, such as Intel x86 Archsim~\cite{intel_archsim}, AMD x86 Simnow~\cite{amd_simnow}, and ARM Fast Models~\cite{arm_fast_models}, can act as GRMs to fuzz processors of different ISAs. }

\red{\noindent\textbf{Support for other coverage metrics.}
We used branch coverage metrics for our study because (i)~they capture the control paths of the design and are crucial in detecting vulnerabilities~\cite{mockus2009test}, and (ii)~existing fuzzers do not capture these paths well (see Section~\ref{sec:advantagesOfHyPFuzz}). 
However, \ourtool{} is not limited to branch coverage metrics. 
\ourtool{} can also target other coverage metrics, such as toggle, finite-state machine~(FSM), condition, expression, etc. To this end, one needs to (i) instrument the DUT for the desired coverage metric and (ii) convert the target coverage metric into a corresponding {\it cover} property that a formal tool can target. 
Commercial hardware tools, such Synopsys \vcs{} or custom compilers~\cite{rfuzz, hur2021difuzzrtl}, already instrument the design for these diverse coverage metrics. 
One can easily generate the \textit{cover} properties for the other coverage metrics (see Appendix~\ref{apd:cov_metric} for further details). 
The rest of the components of \ourtool{}, such as the fuzzer, \textit{scheduler}, \textit{point selector}, and \textit{test case converter} remain unchanged, as they only target the coverage point and not the type of coverage.
Thus, \ourtool{} can be extended to account for other coverage metrics. }

\red{\noindent\textbf{Applicability to other fuzzers.} 
Any coverage-guided hardware fuzzer capable of fuzzing processors, such as~\cite{rfuzz, hur2021difuzzrtl, kandethehuzz, canakci2021directfuzz, ragab_bugsbunny_2022, fuzzhwlikesw}, that outputs the coverage data for each test case can be used as the fuzzer in \ourtool{} as follows. (i)~Modify the \textit{test case generator} to generate the input in the format required by the fuzzer; this is a trivial task since the \textit{test case generator} receives Boolean assignments for all the signals in the DUT for all clock cycles from the formal tool. Since all the data is available, the \textit{test case generator} only needs to filter and format the data. 
(ii)~If the fuzzer uses a coverage metric that is different from branch coverage, then the \textit{property generator} can be modified as  described in Appendix~\ref{apd:cov_metric}.} 

\red{\noindent\textbf{Applicability to other formal tools.}
While we demonstrated \ourtool{} with \JG{}~\cite{jaspergold}, \ourtool{} is agnostic to the underlying formal tool. 
\ourtool{} can use  any formal tool capable of verifying \textit{cover} properties, and thus can use other conventional hardware verification tools like Siemens \questa{}~\cite{questa} and Synopsys \vcformal{}~\cite{vcformal}.}  

\noindent\textbf{FPGA emulations} \blue{is $10\times$ faster than hardware simulation~\cite{hur2021difuzzrtl,rfuzz}.}
\ourtool{} uses hardware simulators to instrument and collect the coverage data. 
Thus, currently, it does not readily support FPGA emulations. 
One can provide FPGA emulation support for \ourtool{} by first instrumenting the DUT using existing tools such as Verific~\cite{verific} or by modifying intermediate representation~(IR) compilers~\cite{rfuzz,hur2021difuzzrtl, canakci2021directfuzz, ragab_bugsbunny_2022}.
The instrumented DUT will then be emulated on the FPGA while the formal tool and the rest of the components of \ourtool{} run on the host machine.
Thus, one can use \ourtool{} to fuzz designs emulated on an FPGA.

%% file: openarch/Conclusion.tex
\section{Conclusion}
Existing hardware fuzzers do not fuzz the hard-to-reach parts of the processor, thereby missing\red{ many} security vulnerabilities. 
\ourtool{} tackles this challenge by using a formal tool to help fuzz the hard-to-reach parts. 
We fuzzed five open-source processors\red{ designs}, including a million-gate \boom{} processor. 
\ourtool{} found three new memory and undefined behavior-related vulnerabilities and detected all the existing vulnerabilities \avebugtimerate{}$\times$ faster than the most recent processor fuzzer. 
It also achieved \aveCovSpeedupvsRandreg{}$\times$ faster coverage than random regression and \aveCovSpeedup{}$\times$ faster coverage than the most recent processor fuzzer.

\noindent\textbf{Responsible disclosure}. We responsibly disclosed the vulnerabilities to the designers.

\section{Acknowledgement}
Our research work was partially funded by the US Office of Naval Research (ONR Award \#N00014-18-1-2058), by Intel's Scalable Assurance Program, and by the European Union (ERC, HYDRANOS, 101055025) \footnote{Views and opinions expressed are however those of the author(s) only and do not necessarily reflect those of the European Union or the European Research Council. Neither the European Union nor the granting authority can be held responsible for them.}. We thank anonymous reviewers and Shepherd for their comments.
Any opinions, findings, conclusions, or recommendations expressed herein are those of the authors and do not necessarily reflect those of the US Government. 

%% file: openarch/Appendix.tex
\appendix
\section*{Appendix} \label{apd:appendix}


\red{\section{New Vulnerabilities Detected by \ourtool{}}\label{apd:NewBugs}}
\red{\noindent\textbf{Vulnerability~\ref{v1}}is in the memory control unit of \cva{} and is similar to out-of bounds memory access vulnerabilities in software programs~\cite{serebryany2012addresssanitizer}. According to the \riscv{} specification~\cite{riscv_home}, a processor must raise an exception when operations try to access data at invalid memory addresses. However, \cva{} will not raise exceptions in the same situation. We detected this vulnerability as a mismatch when \spike{} raised an exception for such operations and interrupted the program, whereas \cva{} continued to execute the program. Because operating systems usually use such exceptions to protect isolated executable memory space, missing them can allow an attacker to access data from all of memory~(CWE-1252~\cite{cwe1252}).}

\red{\noindent\textbf{Vulnerability~\ref{v2}}
is located in the decode stage of \cva{} and is similar to undefined behavior in a software program~\cite{unbehave}. According to the \riscv{} specification~\cite{riscv_home}, the decoder should throw an illegal instruction exception when the destination register~(\texttt{rd}) of instruction \texttt{MULH} is same as either the first~(\texttt{rs1}) or the second~(\texttt{rs2}) source register. This specification will reduce the utilization of multiplier units and increase the performance of processors. It also ensures that a processor will have more available resources for other operations. The vulnerability is that the decoder in \cva{} allows the \texttt{rd} of \texttt{MULH} to share the same register as either \texttt{rs1} or \texttt{rs2}. This means the design of \cva{} violates the ISA specification. We detected this vulnerability when \ourtool{} generated a test case containing \texttt{MULH} with the same \texttt{rd}, \texttt{rs1}, and \texttt{rs2}. \spike{} threw an illegal instruction exception, whereas \cva{} executed the instruction. This vulnerability can result in a potential performance bottleneck when executing applications with heavy multiplier operations, such as machine learning. This vulnerability is similar to the expected behavior violation vulnerability (CWE-440~\cite{cwe440}).}

\red{\noindent\textbf{Vulnerability~\ref{v3}} is a cross-module vulnerability caused by the logic of hardware performance counters~(HPCs) and control state registers~(CSR), and it is similar to undefined behavior in a software program~\cite{unbehave}. HPCs are an important feature in a processor to help programmers find performance bottlenecks in programs~\cite{weaver2013non} and detect malware~\cite{wang2016hardware}. It usually contains multiple physical counters to record hardware behavior during the execution of a program, such as number of instructions executed, number of cycles, number of memory access, number of instruction~(I) or data~(D) cache hit/miss, etc. \cva{} has implemented 14 HPCs for recording various hardware behaviors, and its CSR module is responsible for reading the value of an HPC based on requests from the operating system. However, the reading logic in the CSR module enables access to 32 HPCs, which causes X-propagation when reading nonexistent HPCs. We detected this vulnerability when \spike{} returns regular numbers while \cva{} returns X~(unknown) values. This vulnerability will cause potential issues during synthesis and fail the functions of HPC (CWE-1281~\cite{cwe1281}).}

\section{The \textit{Property Generator} for Different Coverage Metrics}\label{apd:cov_metric}
\ourtool{} currently uses the branch\blue{, condition, and finite-state-machine~(FSM)} coverage metric\blue{s} for evaluation \blue{to demonstrate the compatibility of \ourtool{} with different coverage metrics}.\red{ because this metric is highly related to vulnerability detection~\cite{mockus2009test}.} 
\blue{Our \textit{property generator} is similarly compatible with other coverage metrics also,}
\red{However, the \textit{property generator} of \ourtool{} is also compatible with other coverage metrics, }as we can translate the coverage points to \textit{cover} properties \red{for}\blue{needed by} the formal tools, such as \JG{}~\cite{jaspergold}\red{, to target}.
Usually, the coverage metrics are reported as either a signal name or Boolean expression. Hence, it is always possible to translate these metrics into \systemverilog{} Assertions~(SVA) \textit{cover} properties because the \textit{cover property} is any legal, temporal logic~(TL) expression~\cite{IEEEstd} with the form:
\textit{cover property} $<TL-expression>$, where $<TL-expression>$ can be temporal in general.
We use Listing~\ref{listing:code_cov} as an example to show how to generate \red{an }SVA \textit{cover propert\blue{ies}\red{y}} for different code coverage metrics.




\begin{lstlisting}[language=Verilog, label = {listing:code_cov}, caption={Code snippet showing different coverage metrics.},style=prettyverilog,float,belowskip=-15pt,aboveskip=0pt,firstnumber=1,linewidth=\linewidth]
module code_cov_example (input a,b,c,output d);
    reg [1:0] state_d,state_q;
    always@(*)begin
        case (start_q) begin
            IDLE: begin // FSM
                if (a || b && c) begin // branch, condition
                    state_d = FINISH;
                end else begin
                    state_d = a & c; // expression
                end
            end
            ...
            FINISH: begin // FSM
                d = 1'b1; // toggle
            end
        endcase
    end
endmodule
\end{lstlisting}

\noindent \textbf{Branch coverage} allocates coverage points following the branch statement (i.e., two coverage points, one for \textit{if} branch statement in Line 6 taken and another one for not taken). 
Following the branch statement tree, the conditions of covering the point for taking the \textit{if} branch statement will be $(start\_q == IDLE)$ and $(a |\blue{|} b \&\blue{\&} c)$. And, the condition to check the branch not taken is $(start\_q == IDLE)$ and $not (a |\blue{|} b \&\blue{\&} c)$. The \textit{cover} properties for the two points can be specified in the SVA format as:\\

\textit{cover property} $((start\_q == IDLE) \&\& (a |\blue{|} b \&\blue{\&} c))$ \\
\indent\textit{cover property} $((start\_q == IDLE) \&\& !(a |\blue{|} b \&\blue{\&} c))$\\

\noindent \textbf{Condition coverage} allocates coverage points for all possible combinations of values for the signals in a branch statement (i.e., three 1-bit signals in the \textit{if} branch statement in Line 6 lead to eight condition coverage points.) The \textit{cover} properties for those points can be specified in the SVA format as:\\

\textit{cover property} $(a == 1'b0 \&\& b == 1'b0 \&\& c == 1'b0)$

$\vdots$

\textit{cover property} $(a == 1'b1 \&\& b == 1'b1 \&\& c == 1'b1)$\\

\noindent \textbf{Expression coverage} allocates coverage points for all possible combinations of values for the signals in an assignment (i.e., two 1-bit signals in Line 9 lead to four expression coverage points). The \textit{cover} properties for those points can be specified in the SVA format as:\\

\textit{cover property} $(a == 1'b0 \&\& c == 1'b0)$

$\vdots$

\textit{cover property} $(a == 1'b1 \&\& c == 1'b1)$\\

\noindent \textbf{Finite-state machine~(FSM) coverage} allocates two sets of coverage points to check the FSM in a module. The first set of point checks if state registers have reached all possible state values. The second set of point captures all state transitions in the FSM. Lines 5 and 13 show the two state values and at least one state transition of the $state\_q$ register. The \textit{cover} properties for those points can be specified in\red{ the} SVA format as:\\
(1) checking FSM states:

\textit{cover property} $(state\_q == IDLE)$

$\vdots$

\indent \textit{cover property} $(state\_q == FINISH)$\\
(2) checking state transitions:

\textit{cover property} \\ \indent$(state\_q == IDLE$ \red{$\mapsto$ $\text{eventually }$}\blue{$\#\#1$} $state\_q == FINISH)$\\

\noindent \textbf{Toggle coverage} entails verifying whether specific  bits have been flipped during simulation or not. Line 14 contains a toggle coverage point that verifies if the 1-bit output signal $d$ has been flipped from $0$ to $1$. The \textit{cover} property for this point can be specified in the SVA format as:\\

\textit{cover property} $(d==1'b0 \mapsto d==1'b1)$

\section{Case study example for Boolean assignments and Test case converter}\label{apd:bool_test_eg}

\begin{figure}[!htb]
    \centering
    \includegraphics[trim=15 15 15 15,clip,width=0.85\linewidth]{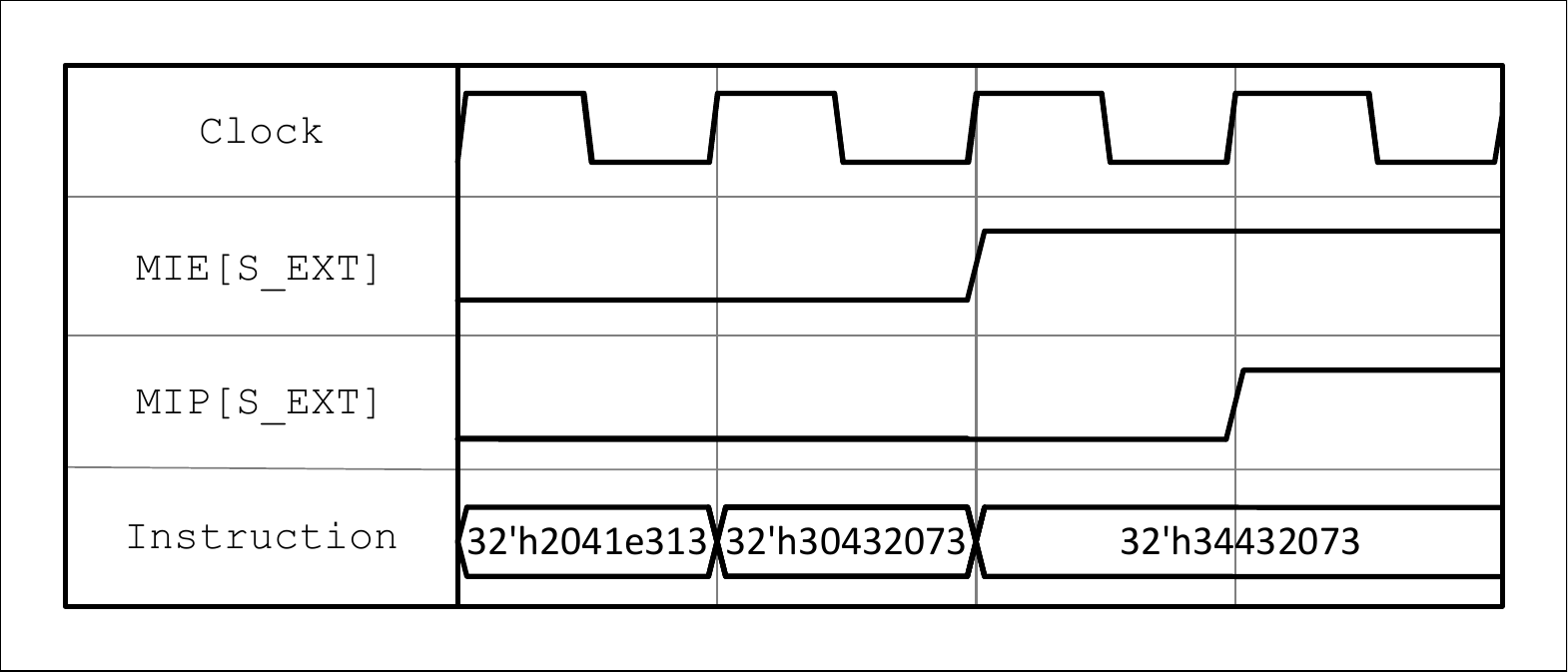}
    \caption{Boolean assignments from \JG{}~\cite{jaspergold}.}
    \label{fig:jg_boolean_assign}
\end{figure}

\begin{figure}[!htb]
    \centering
    \includegraphics[trim=210 80 405 78,clip,width=0.85\columnwidth]{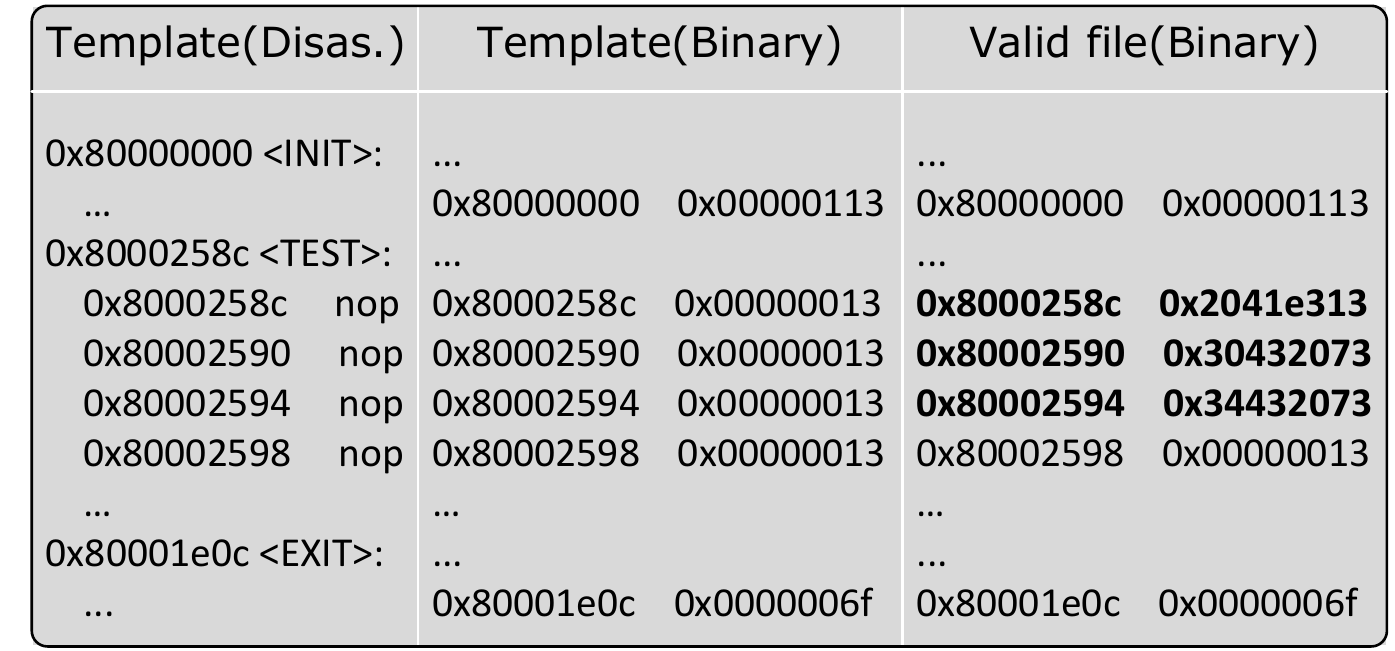}
    \caption{Test case conversion.}
    \label{fig:case_test_case}
\end{figure}

\blue{Figure~\ref{fig:jg_boolean_assign} shows the Boolean assignments output by \JG{} after proving a \textit{cover} property. 
Figure~\ref{fig:case_test_case} shows the template disassembly file used to determine the address of the \textit{TEST} instructions, the binary file template, and the valid binary file generated by the \textit{test case converter} to cover the \texttt{S\_EXT} interrupt branch point. More details are in Section~\ref{sec:advantagesOfHyPFuzz}.}

\section{\moddep{} algorithm}\label{apd:moddep_algo}
\blue{Algorithm~\ref{algo:moddep} is the method to calculate the fanout COI of each module. The module with the highest fanout COI will be selected first. \ourtool{} will prioritize modules based on the dependence before running the experiment. Hence, unlike the \maxuncov{} strategy, the priority of module dependence will not change. More details are discussed in Section~\ref{sec:moddep}.}

\input{Codes/pseudoalgo/moddep}

\section{Evaluation results of $r_{fuzz}$ and $r_{fml}$}\label{apd:eva_rfuzz_fml}
\begin{figure}[!hbt]
    \centering
    \includegraphics[width=\columnwidth]{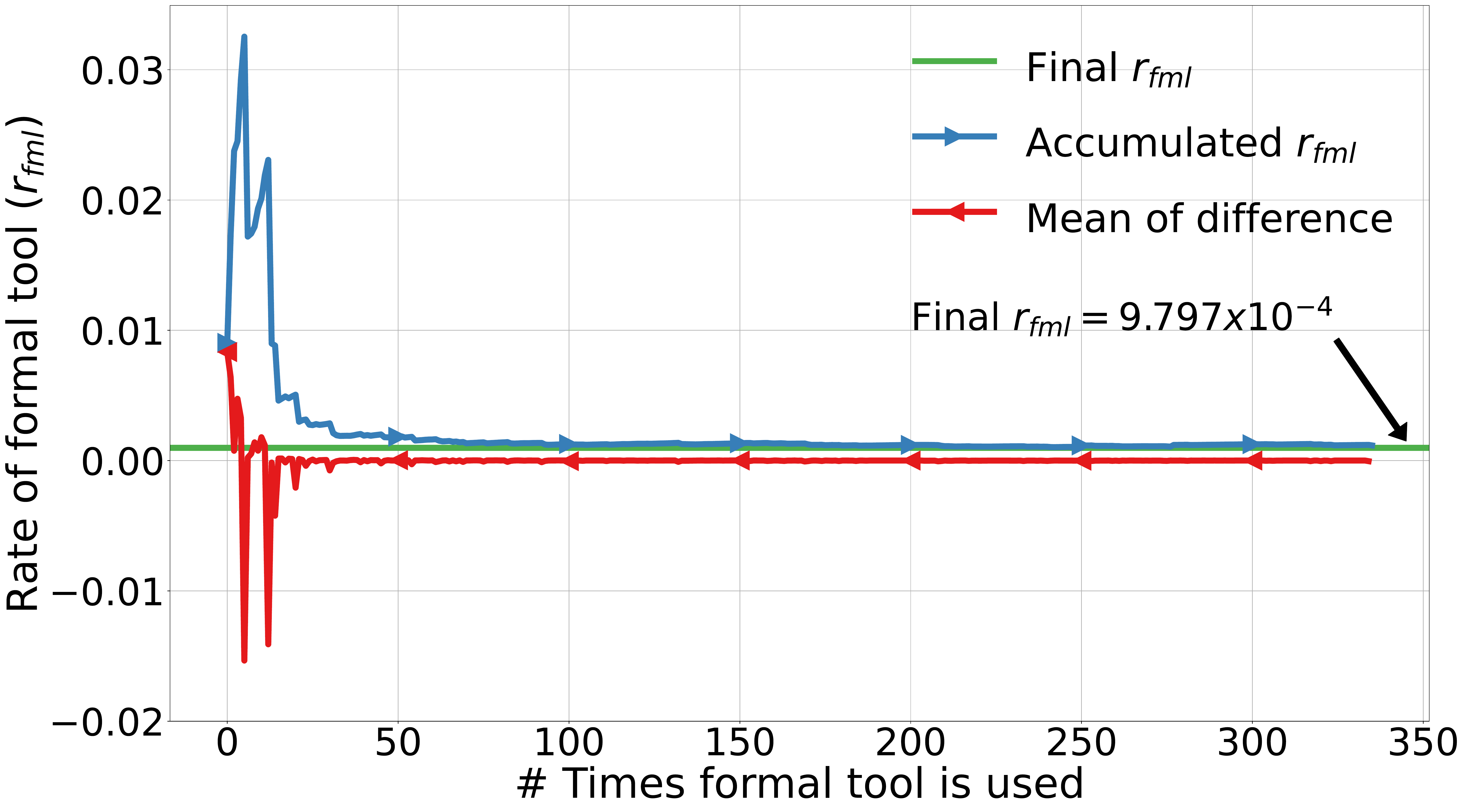}
    \caption{The change of $r_{fml}$ overtime.}
    \label{fig:rfml}
\end{figure}

\begin{figure}[!hbt]
    \centering
    \includegraphics[width=0.85\columnwidth]{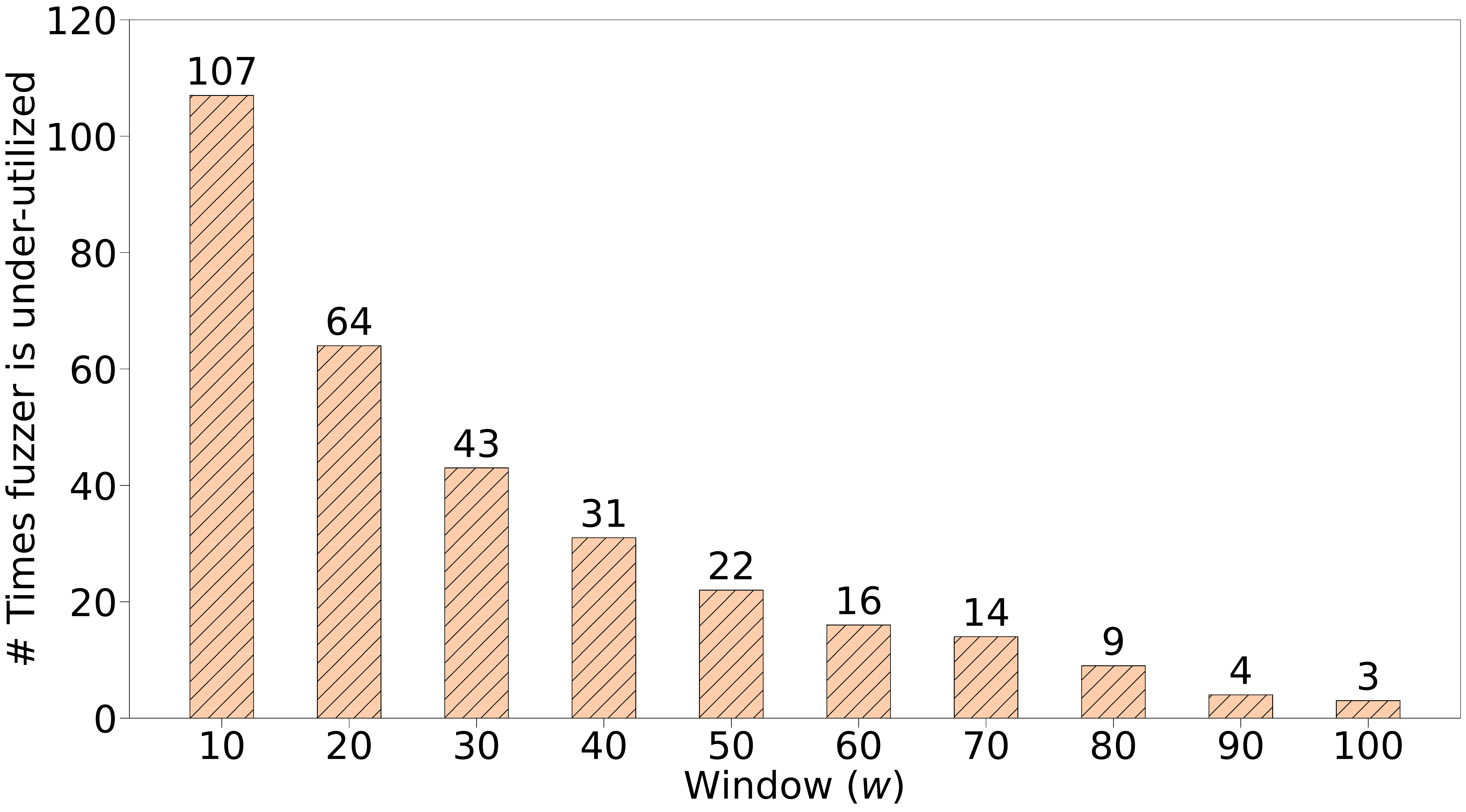}
    \caption{Sampling results of underutilization of fuzzer.}
    \label{fig:rfuzz}
\end{figure}

\blue{Figure~\ref{fig:rfml} and Figure~\ref{fig:rfuzz} show the evaluation results of coverage increment rate of formal tool~($r_{fml}$) and fuzzer~($r_{fuzz}$) respectively on \cva{} processor. Figure~\ref{fig:rfml} evaluates the $r_{fml}$ overtime, and Figure~\ref{fig:rfuzz} evaluates the number of under-utilization of fuzzers when the window size ($w$) is insufficient. More details are discussed in Section~\ref{sec:rateAnalysis}.}

\section{Total branch coverage achieved by random regression, \thehuzz{}~\cite{kandethehuzz}, and \ourtool{}}
\begin{figure*}[htb!]
    \centering
    \begin{subfigure}{0.33\textwidth}
        \includegraphics[width=\textwidth]{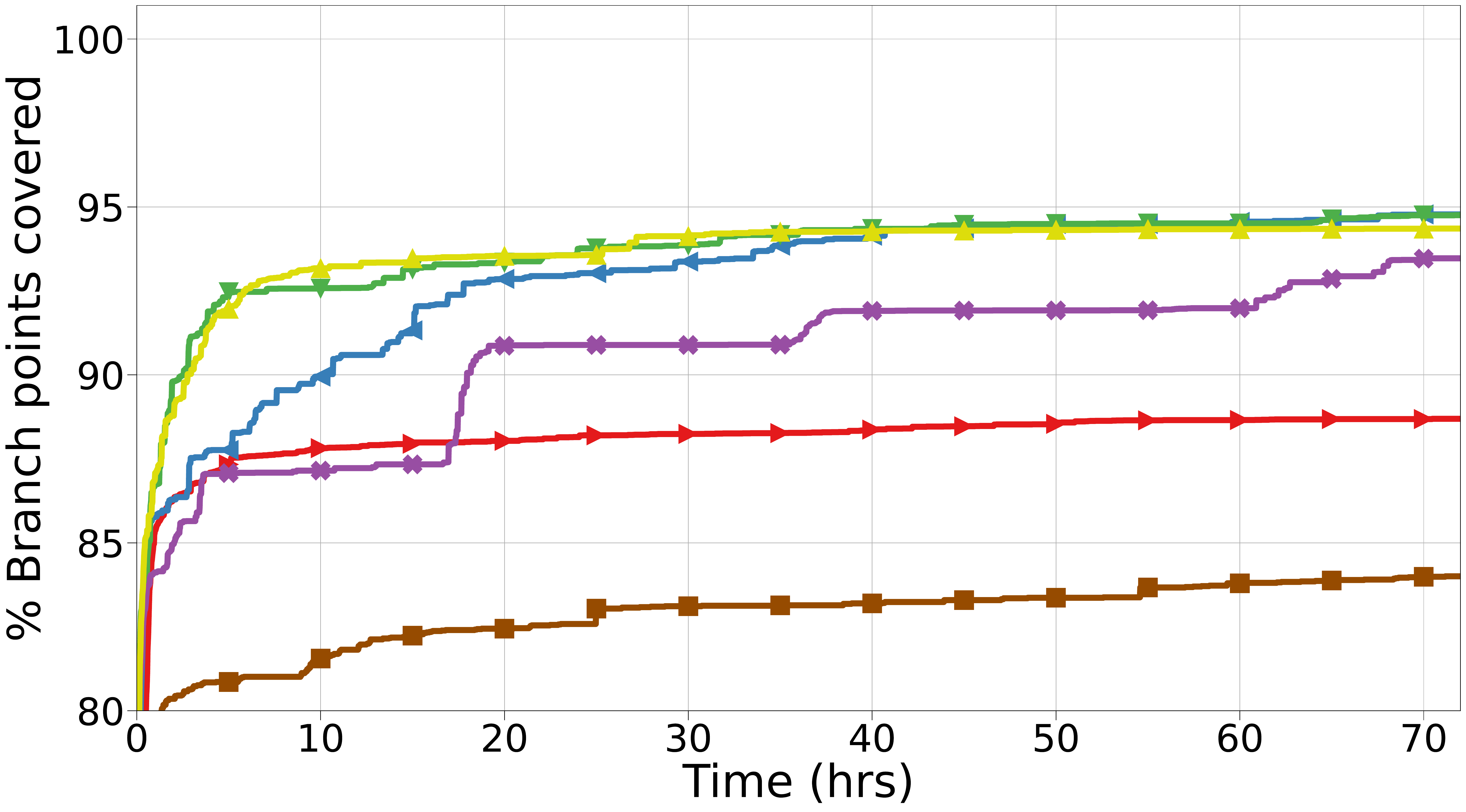}
        \caption{\cva{}~\cite{cva6}}
        \label{fig:cva6_tot_branch}
    \end{subfigure}
    \hfill
    \begin{subfigure}{0.33\textwidth}
        \includegraphics[width=\textwidth]{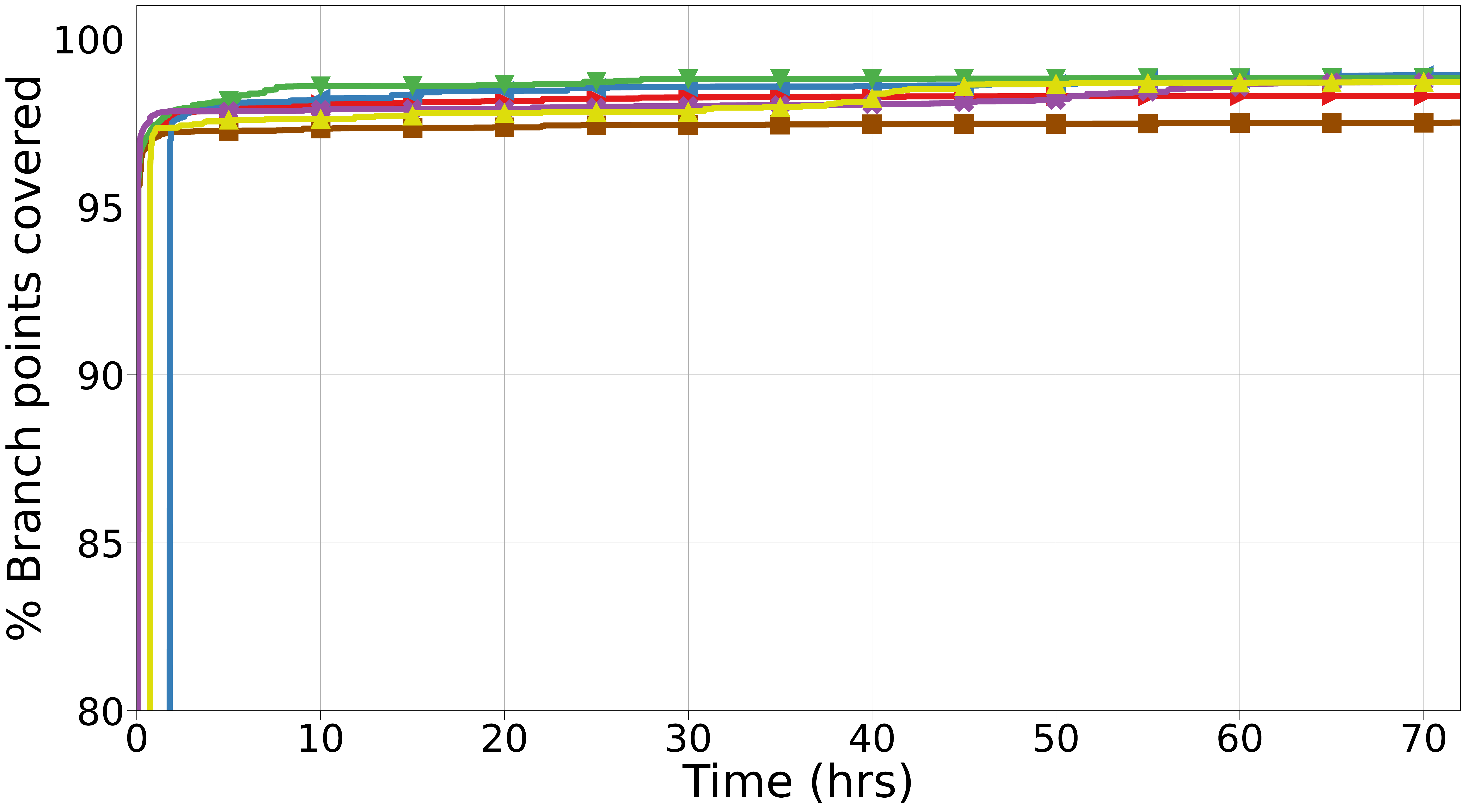}
    \caption{\boom{}~\cite{boom}}
        \label{fig:boom_tot_branch}
    \end{subfigure}
    \hfill
    \begin{subfigure}{0.33\textwidth}
        \includegraphics[width=\textwidth]{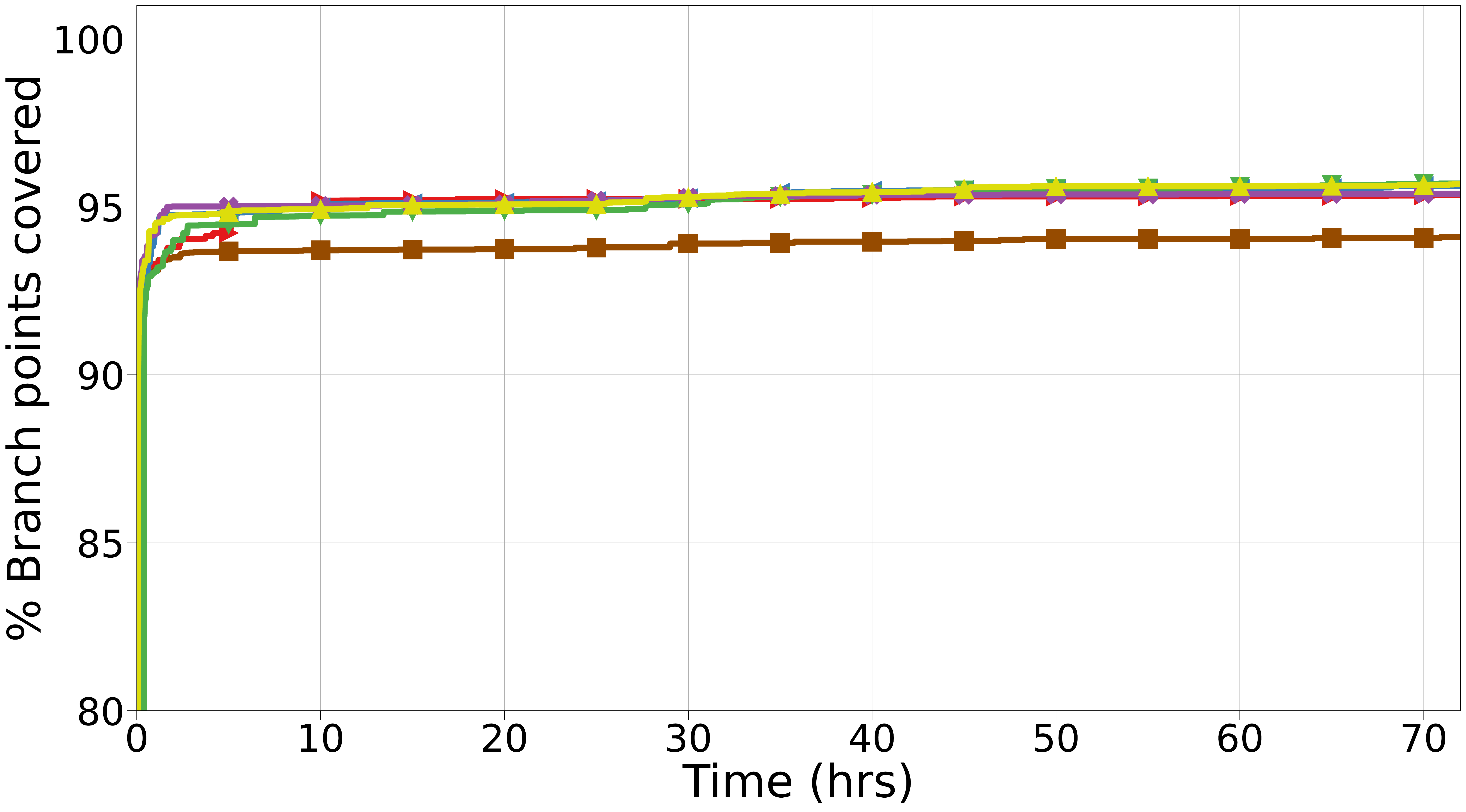}
        \caption{\rc{}~\cite{rocket_chip_generator}}
        \label{fig:rc_tot_branch}
    \end{subfigure}
    \hfill
    \begin{subfigure}{0.33\textwidth}
        \includegraphics[width=\textwidth]{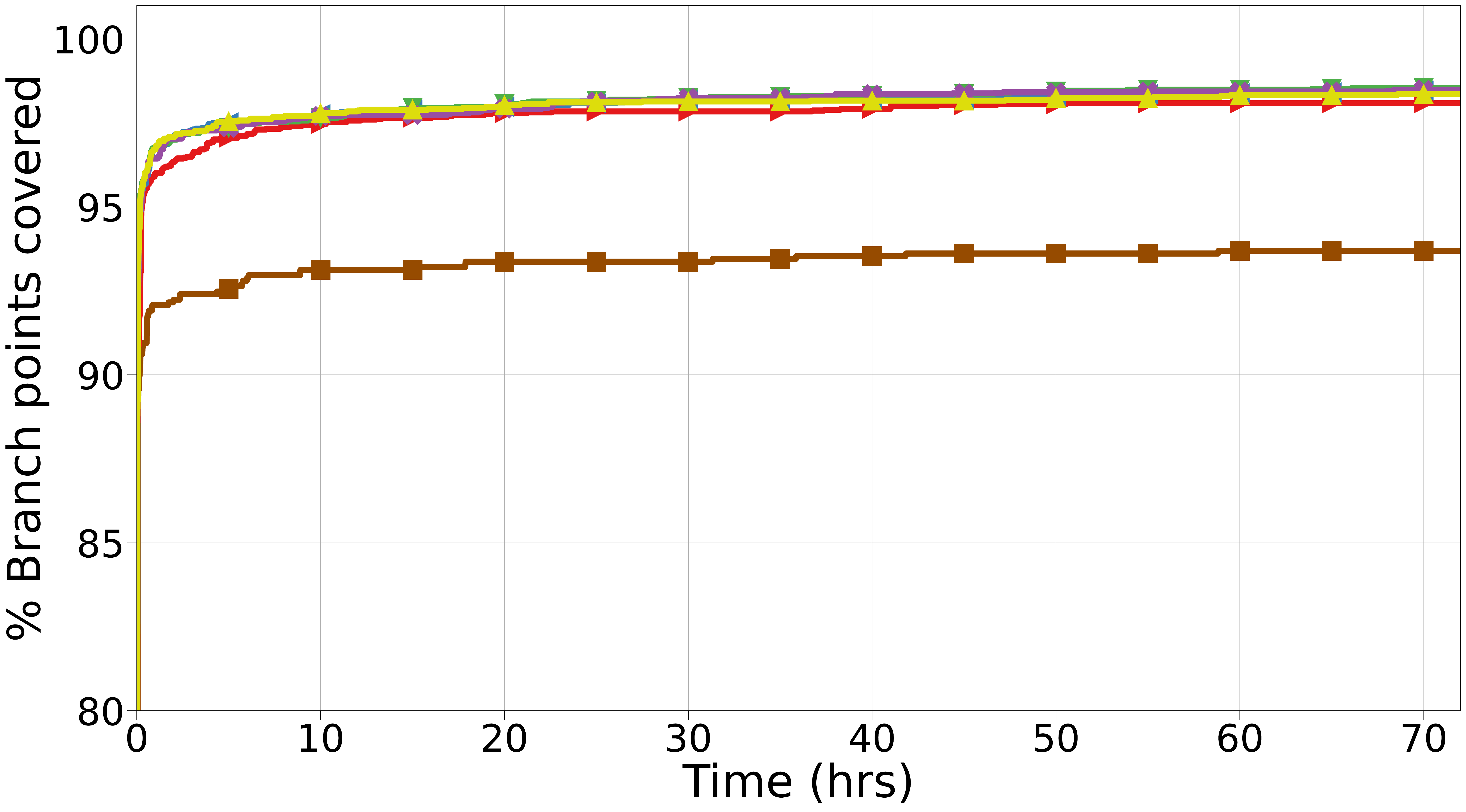}
        \caption{\morkx{}~\cite{mor1kx}}
        \label{fig:mor1kx_tot_branch}
    \end{subfigure}
    \hfill
    \begin{subfigure}{0.33\textwidth}
        \includegraphics[width=\textwidth]{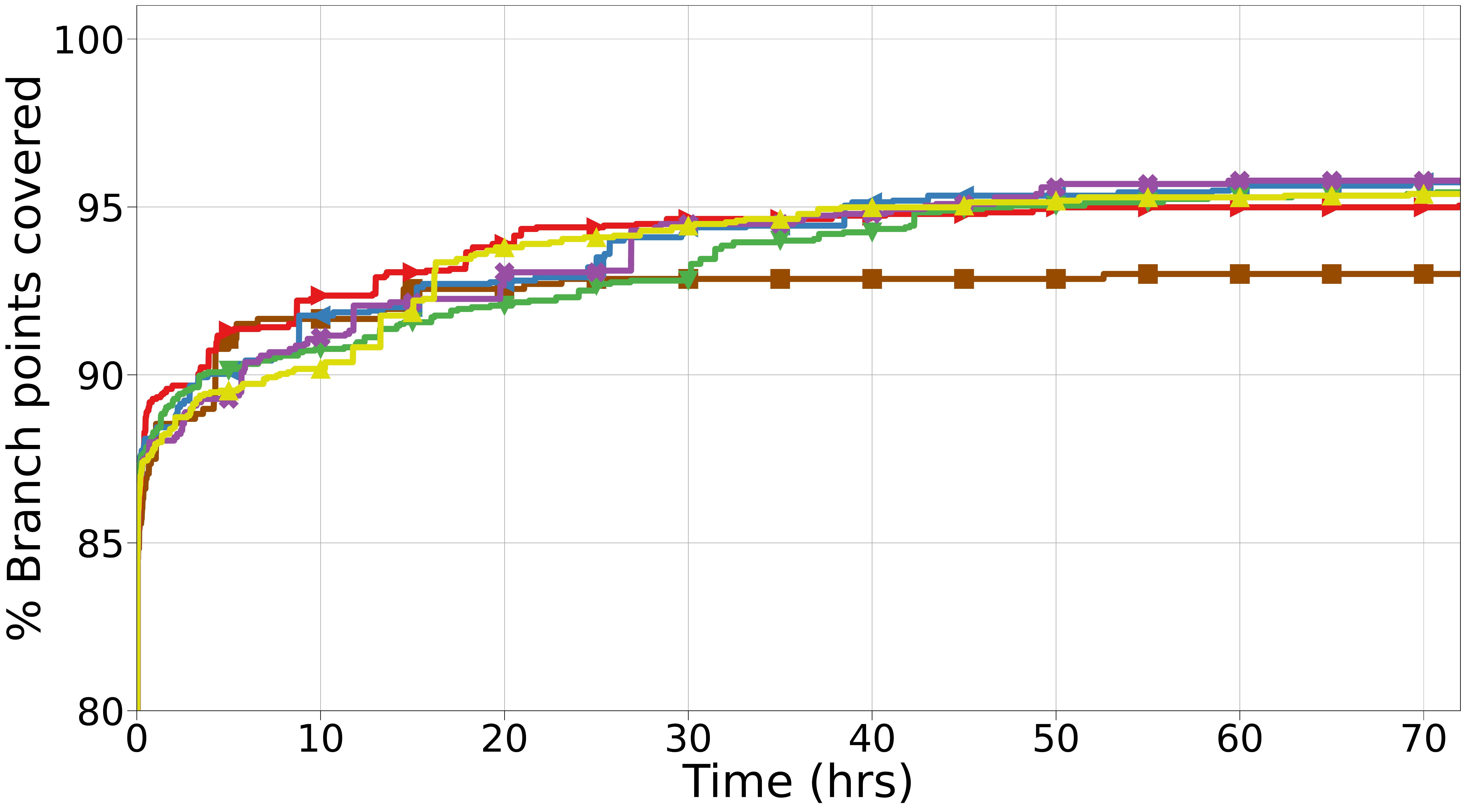}
        \caption{\orth{}~\cite{or1200}}
        \label{fig:or1200_tot_branch}
    \end{subfigure}
    \hfill
    \begin{subfigure}{0.33\textwidth}
        \includegraphics[trim=400 200 200 200,clip,width=\textwidth]{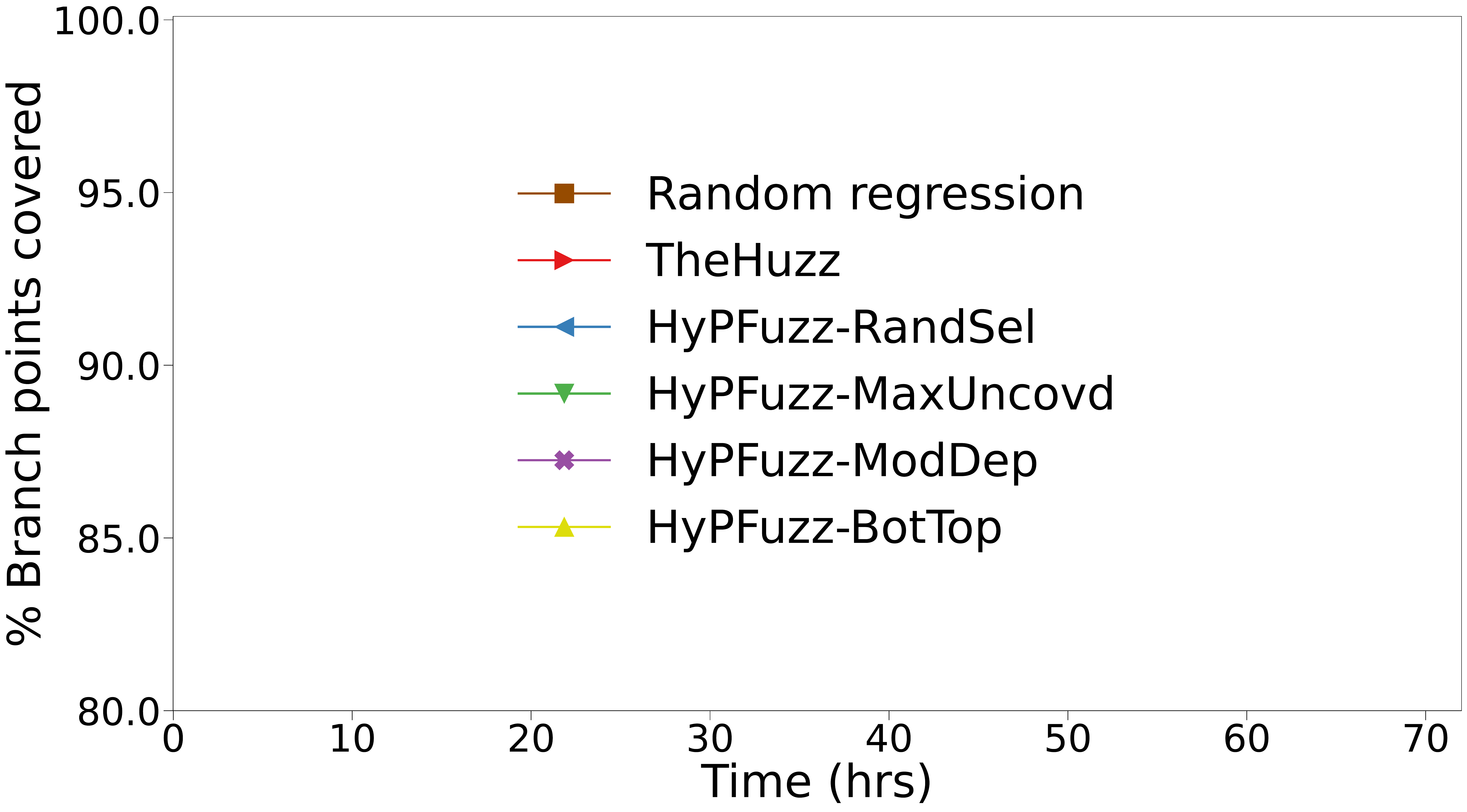}
        \label{fig:legend_tot_branch}
    \end{subfigure}
\caption{Total branch points covered by  random regression, \thehuzz{}~\cite{kandethehuzz}, and \ourtool{}. 
}
\label{fig:tot_branch}
\end{figure*}

\blue{Figure~\ref{fig:tot_branch} shows the branch coverage achieved by random regression, \thehuzz{}~\cite{kandethehuzz}, and different point selection strategies of \ourtool{}. Across the five processors, \ourtool{} achieves \textbf{\avgPerMoreRR{}\%} more coverage than random regression after fuzzing for 72 hours, as seen in Figure~\ref{fig:tot_branch}. 
Also, \ourtool{} achieves \textbf{\avgPerMoreThe{}\%} more coverage than \thehuzz{} after running for the same 72 hours. More details are discussed in Section~\ref{sec:coverageAchieved}.}

%% file: Codes/pseudoalgo/moddep.tex
\begin{algorithm}[!hbt]
\SetFuncSty{textsc}
\SetKwFunction{Output}{Output}
\SetKwFunction{MeasureCOI}{MeasureCOI}
\SetKwFunction{Sort}{Sort}
\SetKwFunction{getOutputs}{getOutputs}
\DontPrintSemicolon
\caption{\moddep{} strategy}\label{algo:moddep}
\KwIn{$M = \{m_0,m_1,...,m_n\}$: a set of modules;}
\KwOut{$M'$, $(|M'|=|M|)$: ordered set of modules based on \moddep{} strategy;  }

$M' \gets \emptyset$\;
$|COI| \gets |M|$ \tcp{store COI of the modules}
\For{$j \gets 0...(|M|-1)$}{
    $coi_j \gets 0$\;
    \For{$output \in $ \getOutputs{$m_j$}}{
        $coi_j \gets coi_j +$ \MeasureCOI{$output$}\;
    }
}

\tcc{sort the module set based on the fanout COI of the modules}
$M' \gets$ \Sort{$M, COI, MaxToMin$}\;
\KwRet{$M'$}\;
\end{algorithm}